\documentclass[10pt,english,aps,amssymb,prb,twocolumn, showpacs, superscriptaddress]{revtex4-2}
\pdfoutput=1
\usepackage[T1]{fontenc}
\usepackage[latin9]{inputenc}
\usepackage{amsmath}
\usepackage{graphicx}
\usepackage{amssymb}
\usepackage{esint}
\usepackage{dsfont}
\usepackage{dsfont}
\usepackage{verbatim}
\usepackage{xcolor}
\usepackage{subcaption}
\usepackage[format=plain]{caption}
\usepackage{natbib}
\usepackage{booktabs}
\usepackage{multirow}

\makeatletter	
\@ifundefined{textcolor}{}
{
 \definecolor{WHITE}{gray}{1}
 \definecolor{RED}{rgb}{1,0,0}
 \definecolor{GREEN}{rgb}{0,1,0}
 \definecolor{BLUE}{rgb}{0,0,1}
 \definecolor{CYAN}{cmyk}{1,0,0,0}
 \definecolor{MAGENTA}{cmyk}{0,1,0,0}
 \definecolor{YELLOW}{cmyk}{0,0,1,0}
 }

\newcommand{\bra}[1]{\langle #1|}
\newcommand{\ket}[1]{|#1\rangle}

\renewcommand{\phi}{\varphi}
\renewcommand{\epsilon}{\varepsilon}

\usepackage{babel}

\begin{document}

\title {Non-Hermitian Topology and Boundary Jordan Chains with Generalized Chiral Symmetry}
\author{Alex Weststr\"om}
\affiliation{Department of Physics, School of Science, Westlake University, Hangzhou 310030, P. R. China}
\author{Wenbu Duan}
\affiliation{School of Physics, Zhejiang University, Hangzhou 310058, China}
\affiliation{Department of Physics, School of Science, Westlake University, Hangzhou 310030, P. R. China}
\author{Jian Li}
\affiliation{Department of Physics, School of Science, Westlake University, Hangzhou 310030, P. R. China}
\date{\today}
\begin{abstract}
We study a generalization of chiral symmetry applicable to non-Hermitian systems and its topological consequences on one-dimensional chains. We find a rich family of topological phases characterized not by a single winding number, but a vector of them. More importantly, we uncover a novel type of bulk-boundary correspondence, where the vector of winding numbers in the bulk corresponds to the set of Jordan chains of various length at the boundary. This in turn leads to highly unconventional chiral-charge distributions on both edges. Our work extends the topological classification of the non-Hermitian AIII class along a new axis.
       
\end{abstract}
\maketitle
\bigskip{}

\section{Introduction}
The classification and understanding of topological phases in condensed matter physics have been central to unraveling the intricate properties of quantum materials. In recent years, non-Hermitian systems have emerged as a promising avenue for exploration, where phenomena not seen in Hermitian systems challenge the conventional wisdom about topological phases. Perhaps the most prominent example of this is the non-Hermitian skin effect and its effect on the concept of bulk-boundary correspondence (BBC) in topological phases \cite{Kunst_2018, Alvarez_2018, Xiong_2018, Yao_2018}.

The spectra of non-Hermitian Hamiltonians, being generally complex with some notable exceptions \cite{Bender_1998, Bender_2007g}, allow for two types of gaps: line gaps and point gaps \cite{Kawabata_2019c}. The topological classification of non-Hermitian systems for point gaps is largely captured in the topological periodic table of Bernard-LeClair classes \cite{Esaki_2011, Gong_2018, Leykam_2017, Lieu_2018, Shen_2018, Zhou_2019} valid for Hamiltonians in any spatial dimension satisfying local symmetries of the form $H = \eta U \mathcal{O}(H) U^\dag$, where $\eta =\pm 1$, $U$ is a unitary matrix acting locally in space, and $\mathcal{O}(H)$ represents the transpose, Hermitian conjugate, or conjugate of $H$. The classification has then mainly been extended by considering non-local $U$ \cite{Liu_2019, De_2019}.


\begin{figure}
	\centering
	\begin{subfigure}[t]{0.22\textwidth}
		\centering
		\includegraphics[width=\linewidth]{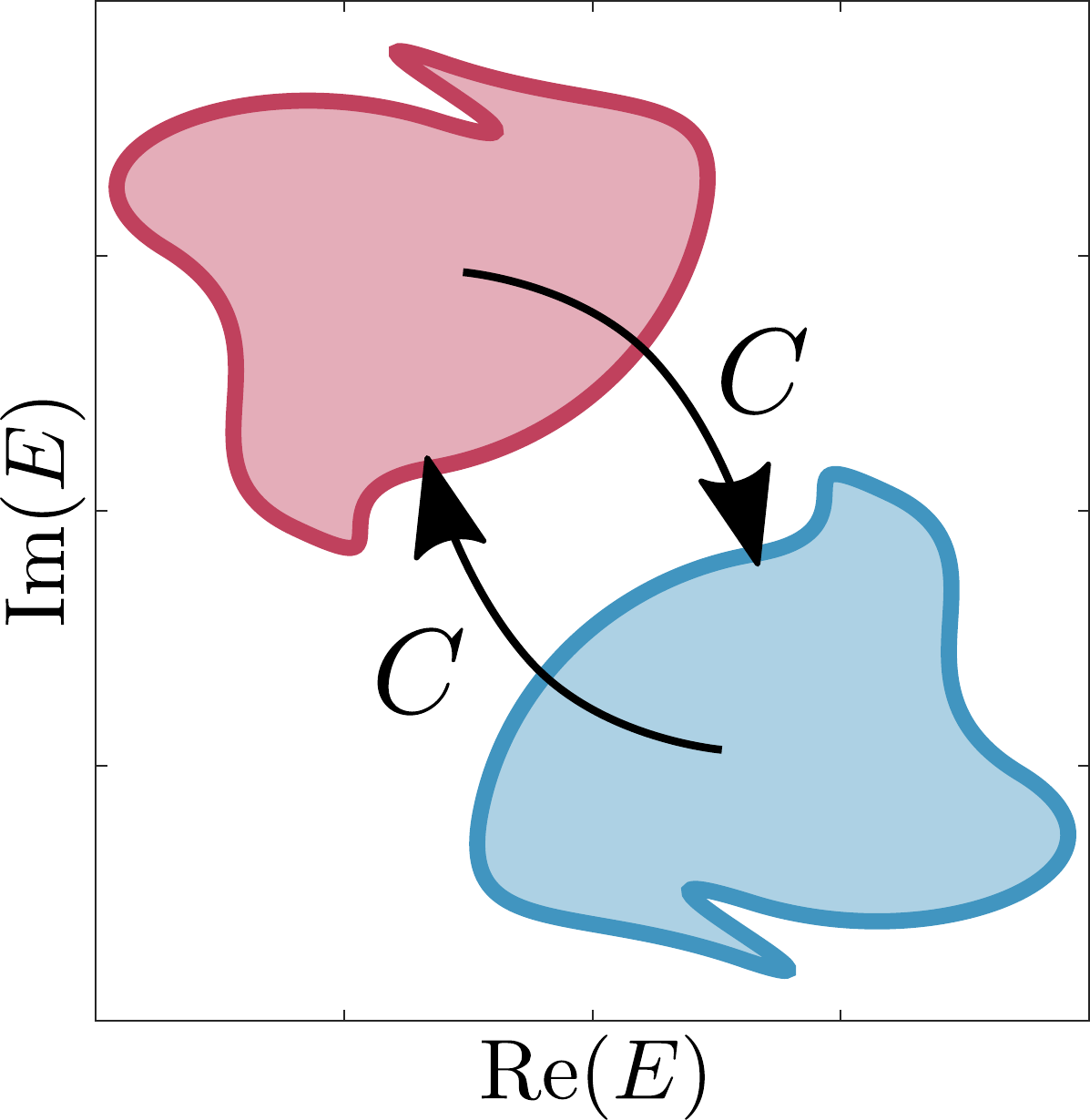}
		\caption{}
	\end{subfigure}
	~
	\begin{subfigure}[t]{0.22\textwidth}
		\centering
		\includegraphics[width=\linewidth]{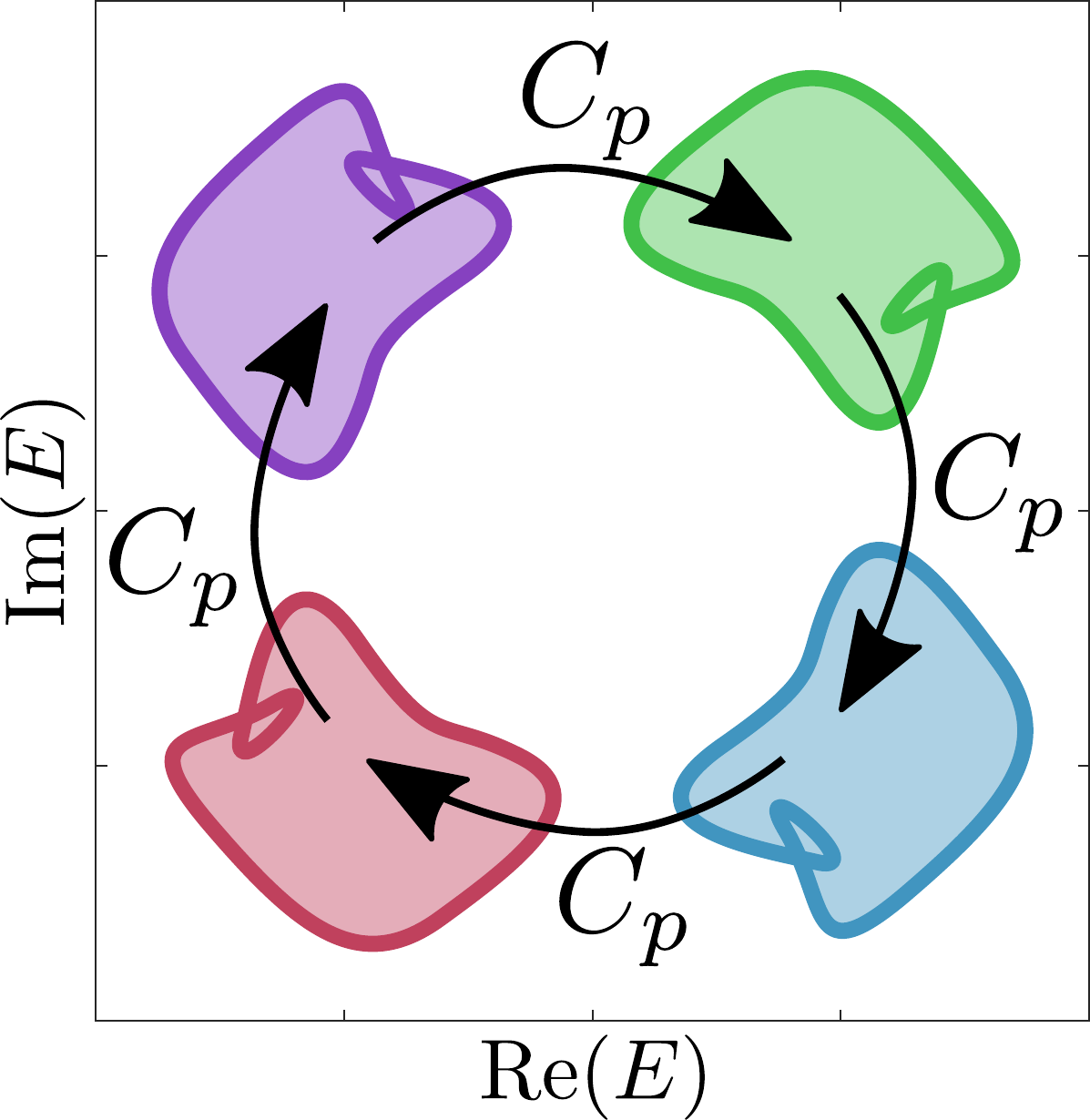}
		\caption{}
	\end{subfigure}
	\caption{Schematic illustrations of complex spectra for non-Hermitian systems with (a) conventional and (b) generalized chiral symmetries. By the notion of $p$-chiral symmetry, the conventional case has $p=2$, while the example in (b) has $p=4$. In both cases, the spectra consist of sectors that are mapped cyclically into one another by the $p$-chiral operator $C_p$.}\label{fig:schemspec}
\end{figure}	

A hitherto less explored extension is allowing $\eta$ to take on other values on the unit circle than just $\pm 1$. One example of such an extension is a generalized chiral symmetry (see Fig.~\ref{fig:schemspec}) as introduced for non-interacting fermions in \cite{Marques_2022}, where a generalization of Lieb's theorem applicable to multipartite systems has been proved. This generalized symmetry was subsequently used to construct higher-root topological insulators \cite{Viedma_2023}. In addition, generalized chiral symmetry is also present in Baxter's clock model for parafermions \cite{Marques_2022, Baxter_1989, Alicea_2016}, and a more general version of this symmetry was also shown to allow for the existence of higher-order topological phases in Hermitian systems \cite{Ni_2019, Li_2021}.

In the interest of extending the topological classification of non-Hermitian systems, we dedicate this work to studying the rich topological properties of one-dimensional chains possessing this generalized chiral symmetry. We will first provide a brief introduction to the generalized chiral symmetry and then define the relevant topological phases with the associated topological invariant. Building on that, we will present a novel BBC and discuss how it differs from conventional chiral systems. Finally, we will summarize our findings and discuss some potential future research directions.


\section{$p$-Chiral Symmetry}\label{sec:sympre}

\begin{figure*}
	\centering
	\begin{subfigure}[t]{0.32\textwidth}
		\centering
		\includegraphics[width=\linewidth]{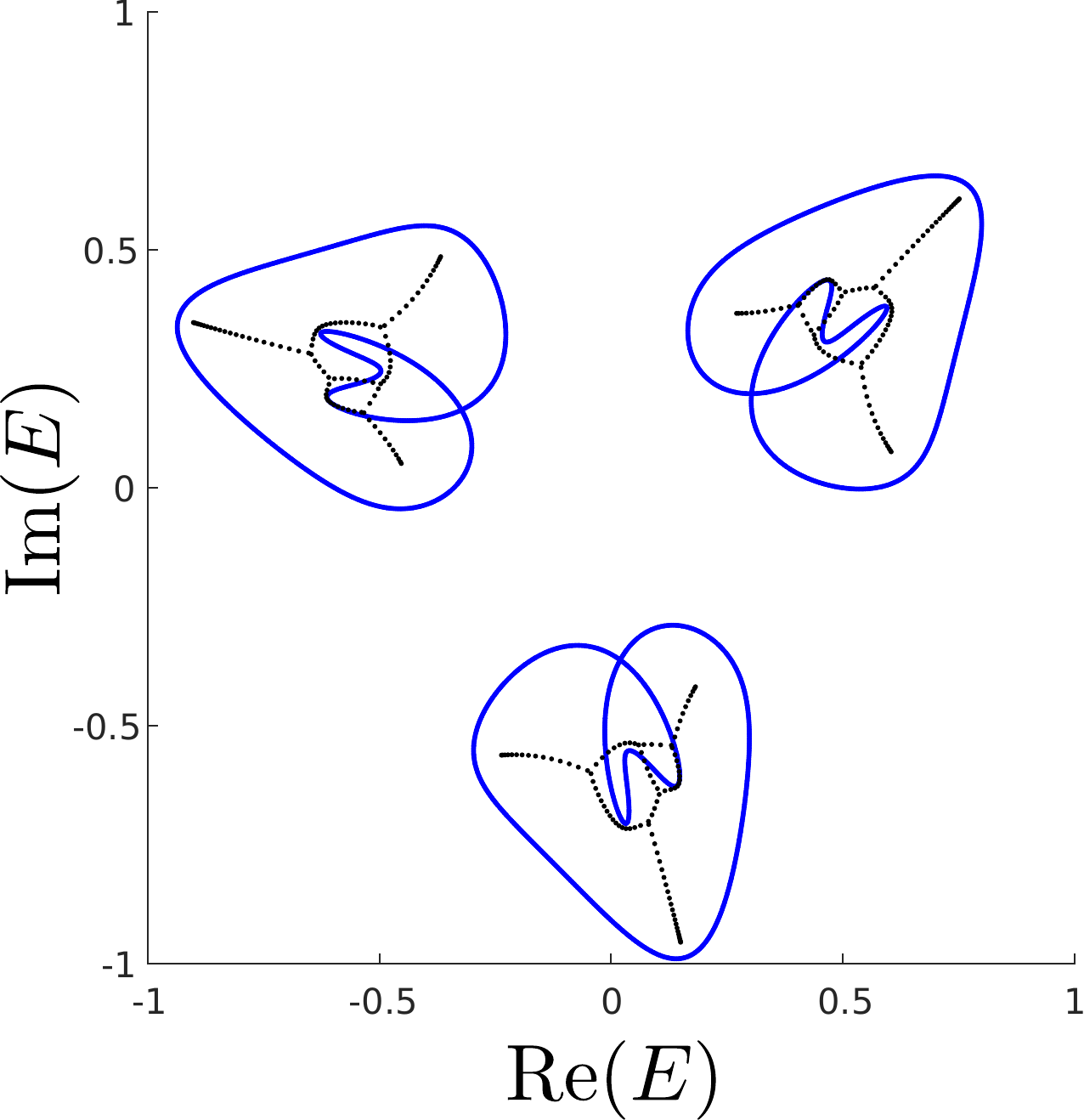}
		\caption{}
	\end{subfigure}
	~
	\begin{subfigure}[t]{0.32\textwidth}
		\centering
		\includegraphics[width=\linewidth]{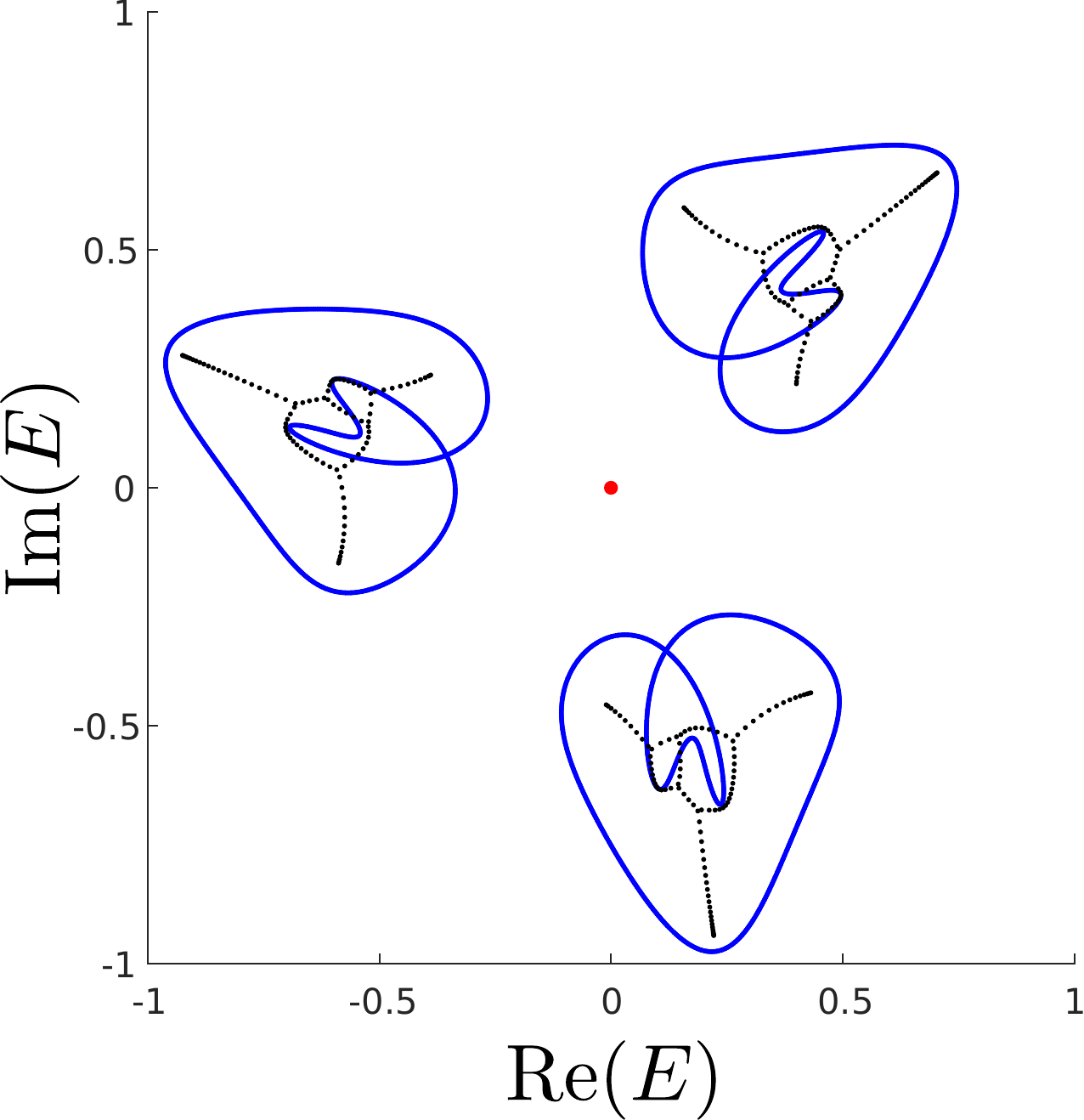}
		\caption{}
	\end{subfigure}
	~
	\begin{subfigure}[t]{0.32\textwidth}
		\centering
		\includegraphics[width=\linewidth]{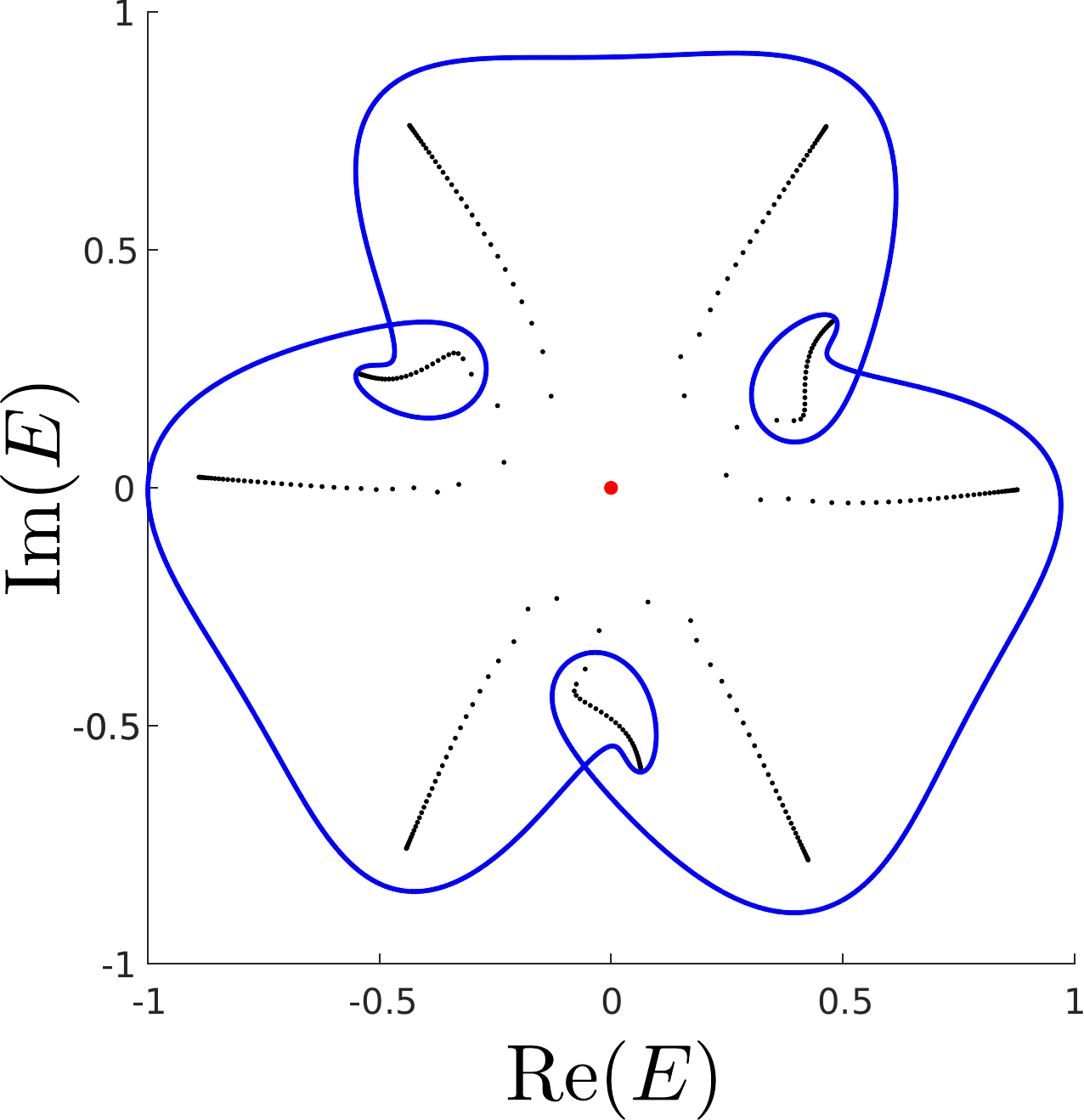}
		\caption{}
	\end{subfigure}
	\caption{(a) An example of the spectrum for a $p$CS system with $p=3$ in the trivial phase. The blue lines and the black points are the periodic- and open-boundary spectra, respectively. The open-boundary spectrum is calculated for a system with 100 sites. The energies have been rescaled such that the largest energy modulus is set to one. (b) Same as (a) except for a non-trivial phase, where $\mathbf{W} = (1, 2, -3)$. The zero-energy eigenvalue of the open-boundary spectrum is colored in red. (c) Similar to (a) and (b) except for a system with nonzero total winding $\mathbf{W} = (1, 1, -3)$. The toy models used in this figure are randomly generated by setting $a(k) = c + d e^{i n k}$ with random complex numbers $c$ and $d$ ($\vert c\vert > \vert d\vert$ for trivial phases; $\vert c\vert < \vert d\vert$ for topological phases) chosen differently for each block and $n$ determined by the components of $\mathbf{W}$.}\label{fig:spectra}
\end{figure*}

In its standard form, chiral symmetry is expressed as an anticommutation relation $\{H, C\} = 0$ between a Hamiltonian $H$ and a unitary operator $C$ with the property $C^2 = 1$. The generalized chiral symmetry -- from hereon referred to as $p$-chiral symmetry ($p$CS) -- replaces the anticommutation relation by
\begin{equation}\label{eq:pcs}
	C_p H C_p^\dag = \omega_p H,
\end{equation}
where the $p$-chiral operator satisfies $(C_p)^p = 1$ with $p$ a positive integer and $\omega_p = e^{\frac{2\pi i}{p}}$ the $p$-th root of $1$. In this notation, the standard chiral symmetry corresponds to $p = 2$. When $p>2$, a non-trivial Hamiltonian is necessarily non-Hermitian. An immediate consequence of $p$CS is that the spectrum will have a rotational symmetry in the complex energy plane: for any (right) eigenvector $\psi$ of the Hamiltonian with eigenvalue $\varepsilon$, $C_p\psi$ will also be a (right) eigenvector with eigenvalue $\omega_p^{-1} \varepsilon$. In fact, repeatedly applying $C_p$ on $\psi$ gives us a $p$-tuple of eigenvectors whose energies are related by different powers of $\omega_p$, as illustrated in Fig.~\ref{fig:schemspec}.

Since our focus will be on \textit{gapped} topological phases (with respect to zero energy), we will assume $C_p$ in its diagonalized basis to be
\begin{equation}
	C_p = \text{diag}\left[1, \omega_p,\ldots, \omega_p^{p-1}\right]\otimes \mathbb{I}_{N\times N},
\end{equation}
where $N$ is the number of additional degrees of freedom in each chiral sector. Any other choice of sizes for the different chiral sectors would lead to zero-energy bands as per Lieb's theorem \cite{Marques_2022, Gohsrich_2024}. In the same basis, a generic Hamiltonian satisfying Eq.~\eqref{eq:pcs} must be of the form
\begin{equation}\label{eq:pcham}
		H = 
		\begin{pmatrix}
			0 & 0 & \cdots & 0 & \mathbf{a}_{1,p}\\
			\mathbf{a}_{2,1} & 0 & \cdots & \cdots & 0\\
			0 & \mathbf{a}_{3,2} & \ddots &  & \vdots\\
			\vdots & \ddots & \ddots & \ddots & \vdots\\
			0 & \cdots & 0 & \mathbf{a}_{p,p-1} & 0
		\end{pmatrix},
\end{equation}
where each $\mathbf{a}_{j+1,j}$ ($p+1$ is identified with $1$) is an $N$-by-$N$ matrix. Similarly to the standard chiral symmetry, this form allows us to make a sublattice interpretation, but instead of having only two sublattices, we now have $p$ sublattices that are cyclically connected by the Hamiltonian.

The spectrum of the Hamiltonian in Eq.~\eqref{eq:pcham} can be solved from the characteristic equation
\begin{equation}
	\det\left(\lambda^p - \mathbf{a}_{p,p-1}\cdots\mathbf{a}_{2, 1}\mathbf{a}_{1,p}\right) = 0,
\end{equation}
where we explicitly see the aforementioned rotational symmetry manifest: if $\varepsilon$ is a solution to the characteristic equation, then so is $\omega_p^n \varepsilon$ for any integer $n$ because $(\omega_p^n)^p=1$. The above expression also makes it evident that all individual $\mathbf{a}_{j+1,j}$ need to be nonsingular for $H$ to be nonsingular. Throughout this paper, addition with respect to this sublattice-like index $j$ will be taken as modulo $p$ unless otherwise stated.

The topological nature of zero-energy edge modes in $p$CS systems can now be understood through the following argument: an isolated zero-energy state in a $p$CS system must due to symmetry be an eigenvector of the chiral operator with an associated eigenvalue which we will refer to as \textit{chiral flavor} (in the literature, ``flavor'' is typically called ``charge'', but we reserve this term for referring to the amount -- a concept that will be clarified later in the text -- of a given flavor). This state cannot be lifted from zero energy without breaking the symmetry, closing the gap, or combining it with zero-energy states of complementary chiral flavors. The last option follows from the fact that a state with nonzero energy $\varepsilon$ must have $p-1$ partner states with energies $\omega_p^n \varepsilon$, $n = 1, 2, \ldots, p-1$ that cyclically map into each other via repeated applications of the Hamiltonian. For finite-energy states, this implies a nonzero component for all chiral flavors, further implying that all flavors have to be present when removing states from zero energy. The above reasoning already hints at the first novel feature of $p$CS topology: \textit{there will not necessarily be an equal number of zero modes at each end of a finite chain.}

\section{Winding and Topology}\label{sec:windtop}
Unlike Hermitian physics, the spectrum for a non-Hermitian system with open boundary conditions (OBC) can be vastly different from that of the corresponding one with periodic boundary conditions (PBC) \cite{Yao_2018}. As it pertains to the setups studied in this paper, one can define a spectral winding number (or vorticity) \cite{Gong_2018, Shen_2018} 
\begin{equation}\label{eq:winding}
	\mathcal{W}_\varepsilon(\mathbf{b}_k) = \frac{1}{2\pi i}\int_{-\pi}^\pi \text{Tr}\left[\left(\mathbf{b}_k - \varepsilon\right)^{-1}\frac{\partial \mathbf{b}_k}{\partial k}\right]dk
\end{equation}
for a continuous and periodic matrix function $\mathbf{b}_k: S^1\to \text{GL}(n, \mathbb{C})$, where $S^1$ is the circle and $\text{GL}(n, \mathbb{C})$ is the set of complex-valued $n\times n$ matrices. This winding number can be calculated for any point $\varepsilon$ in the complex energy plane, provided it does not coincide with any of the eigenvalues of $\mathbf{b}_k$ for any $k$. The winding number tells us how many times the spectrum winds around $\varepsilon$ as we go around the Brillouin zone -- if it is nonzero, there will be skin modes present at that energy \cite{Borgnia_2020, Okuma_2020, Zhang_2022} in a semi-infinite chain \cite{windingnote}. To ensure that our non-Hermitian system remains gapped even when skin modes are taken into account, it is imperative that we restrict ourselves to systems where the winding number around $\varepsilon=0$ is zero.

The winding number for a Hamiltonian of the form Eq.~\eqref{eq:pcham}, as shown in App.~\ref{sup:wind}, is given by the sum of the individual winding numbers of each $\mathbf{a}_{j+1,j}(k)$. The topological phase for $p$CS systems with a vanishing total winding number is then uniquely determined by $p-1$ individual winding numbers. Ignoring this redundancy, the topological invariant now becomes a vector
\begin{align}\label{eq:vectorinv}
	\mathbf{W} &\equiv
	\bigl(
		\mathcal{W}_0(\mathbf{a}_{2,1}), \mathcal{W}_0(\mathbf{a}_{3,2}), \ldots, \mathcal{W}_0(\mathbf{a}_{p,p-1}), \mathcal{W}_0(\mathbf{a}_{1,p})
	\bigr) \nonumber\\
        &\equiv (W_1, W_2, \ldots, W_{p-1}, W_p).
\end{align}
We remark that in the conventional $p=2$ Hermitian chiral class, the values of $\mathcal{W}_0(\mathbf{a}_{2,1})$ and $\mathcal{W}_0(\mathbf{a}_{1,2})$ are locked by Hermiticity to be exactly opposite. This is why the topological classification is reduced from $\mathbb{Z}^2$ to $\mathbb{Z}$ for the conventional AIII symmetry class when moving from non-Hermitian to Hermitian settings.

The zero-total-winding constraint implies that any such phase transition requires a change of at least two individual winding numbers. In other words, zero-total-winding phases are in general separated by intermediate gapless phases, where, again, the gaplessness is defined in terms of the presence of skin modes. This means that in a system where the zero-total-winding spectrum consists of separate islands, as exemplified in Fig.~\ref{fig:spectra}(a) and (b), the phase transition is accompanied by transitions in the topology of the spectrum as well: in the intermediate phase where the overall winding is nonzero, there must be at least one spectral band wrapping around zero energy \cite{Nehra_2021, Nehra_2022, Gohsrich_2024}, as in Fig.~\ref{fig:spectra}(c); the transition between separate island bands and such a single wrapping band happens when the islands meet and reconnect at zero energy. Note that it is also possible to construct models where the spectrum wraps around zero energy but the total winding remains zero, making the fact that we are dealing with point-gap spectra all the more obvious. We present an example of this in App.~\ref{sup:toy}.

As a last point of this section, we mention that in the case of a composite $p = mn$ for some integers $m$ and $n$, one can consider breaking $p$CS into a lower $m$CS by for example introducing additional matrix elements between consecutive cosets (say, $r+1$ and $r$) of flavors originally labeled by $\{r + lm\}_{l=0,1,\ldots,n-1}$ with $0\leq r < m$. The new chiral operator, $C_p^n$, does not distinguish between the original $p$-chiral flavors belonging to the same coset labeled by $r$. If we break the $p$CS down to $m$CS without closing the gap, the new vector invariant $\mathbf{W}^{(m)}$ is related to the original one $\mathbf{W}^{(p)}$ by the coset summation: $W^{(m)}_r = \sum_l W^{(p)}_{r+lm}$.

\section{Bulk-Boundary Correspondence}\label{sec:bbc}
Conventionally, there is an immediate relationship between the bulk topological invariant derived from a PBC calculation and the boundary modes of the corresponding OBC system: the magnitude of the topological invariant (for which zero corresponds to the trivial phase) equals the number of topological boundary modes at both ends of the open chain; the sign of the topological invariant registers a sense of orientation for the boundary modes. Now, with a vector of invariants describing the overall topological phase of our system, the conventional BBC needs to be revised.

\begin{figure}[t]
	\includegraphics[width = 0.95\linewidth]{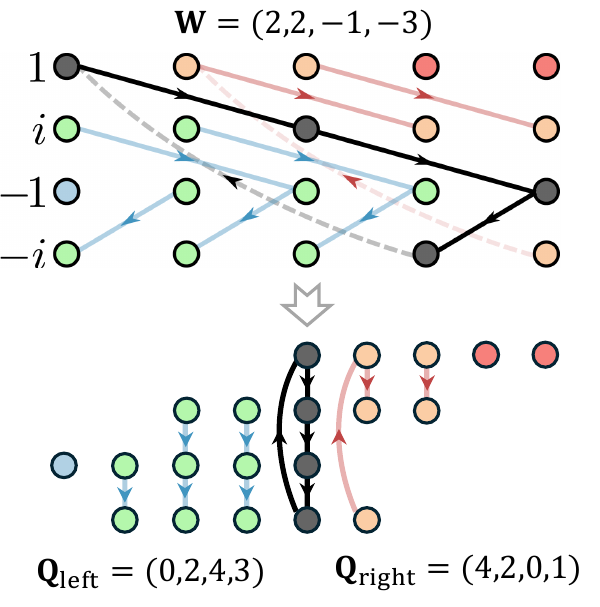}
	\caption{Illustration of a sweet-spot model with the vector of winding numbers being $\mathbf{W} = (2, 2, -1, -3)$. Here, each column of dots corresponds to a site in real space and each row corresponds to a chiral flavor $\{\omega_4^j\}_{j=0,1,2,3}$ labeled on the left. The black dots are part of a closed loop which, when replicated, makes up the bulk of the chain; the blue (red) arrows connecting green (orange) dots correspond to incomplete loops, i.e. Jordan chains, on the left (right) edge; the blue and red dots correspond to conventional zero modes which can be regarded as Jordan chains of length $1$. The lower part of this figure is a straightened-up version of the original one in the upper part. The composition of the boundary Jordan chains is determined by the winding number vector $\mathbf{W}$, defining a new type of bulk-boundary correspondence.}\label{fig:sweetspot}
\end{figure}

A heuristic approach to understand the relationship between the vector invariant and the boundary behavior is to consider a so-called ``sweet spot'' model for which the PBC version merely amounts to
\begin{equation}
	a_{j+1,j}(k) = e^{i n_j k},
\end{equation}
where $n_{j=1,\ldots, p} \in \mathbb{Z}$ are the winding numbers, and thus the zero-total-winding requirement simply means $\sum_j n_j = 0$. In real space, $e^{i n k}$ translates to a hopping term of length $\vert n\vert$ with hopping direction being determined by $\text{sgn}(n)$; the step-like pattern of the matrix elements $a_{j+1,j}$ implies that the hoppings traverse the $p$ chiral flavors in a directed and periodic fashion. As illustrated in Fig.~\ref{fig:sweetspot}, the zero-total-winding condition hence implies that when we draw the hopping terms between different sites as directed links, they form closed loops in the bulk, as well as possibly open chains (including isolated sites) at the boundary.

As a concrete example, let us consider a system with $\mathbf{W} = (2, 2, -1, -3)$. We will see that this example captures the new features not present in conventional chiral systems. If we now collect the number of sites $c_n$ that are not part of a closed loop at the left (right) boundary for each flavor labeled $n$ into a vector $\mathbf{Q}_\text{left (right)} = (c_1, c_2, \ldots, c_p)$, Fig.~\ref{fig:sweetspot} gives us $\mathbf{Q}_\text{left} = (0, 2, 4, 3)$, and $\mathbf{Q}_\text{right} = (4, 2, 0, 1)$ for our particular example, which appears to be at odds with $\mathbf{W}$ according to the conventional BBC. Notably, the largest number in both $\mathbf{Q}_\text{left}$ and $\mathbf{Q}_\text{right}$ even exceeds the winding number with the maximum modulus in $\mathbf{W}$. To resolve this puzzle, we have to shift our attention from boundary sites to boundary chains. An open chain formed by $l$ sites represents a Jordan block of dimension $l$ and hence by definition is a Jordan chain consisting of $l$ linearly independent vectors in the Hilbert space. Here, an isolated site is a special case with $l=1$. Counting Jordan chains at the boundary and grouping them according to the flavors of their ends, we obtain a sequence $\mathbf{C} = (2, 2, -1, -3)$, where the positive/negative sign indicates Jordan chains on the right/left edge. If we group the Jordan chains according to their starting flavors instead, we obtain a sequence $\mathbf{C}' = (3, -2, -2, 1)$, which is clearly equivalent to $\mathbf{C}$ up to a change of labeling and sign conventions as long as the cyclic order is maintained. Note that a specific flavor cannot accommodate ends (nor beginnings) of Jordan chains at both edges because of the unidirectional hopping.

The coincidence between $\mathbf{C}$ and $\mathbf{W}$ reveals a new type of BBC: \textit{the vector of bulk topological invariants corresponds to the collection of boundary Jordan chains}. This is a peculiar consequence of $p$CS and also the main result of this paper. In fact, the boundary data $\mathbf{C}$, and thus $\mathbf{W}$, completely determine the structure of the boundary Hilbert space encoded in $\mathbf{Q}_\text{left}$ and $\mathbf{Q}_\text{right}$, as the latter numbers can be directly computed as follows (see App.~\ref{sup:algo} for the derivation):
 \begin{enumerate}
 	\item Define a cumulative sum vector $\mathbf{Q}$ with its component $(\mathbf{Q})_j \equiv \sum_{j'=1}^{j-1} W_{j'}$. For example, $\mathbf{Q} = (0, 2, 4, 3)$ when $\mathbf{W} = (2, 2, -1, -3)$.
 	\item $\mathbf{Q}_\text{left} = -\min \mathbf{Q} + \mathbf{Q}$.
 	\item $\mathbf{Q}_\text{right} = \max \mathbf{Q} - \mathbf{Q}$.
 \end{enumerate}
Note that this is valid not only for the sweet-spot model, but also in the general case as is explained in App.~\ref{sup:algo}. In general, $\mathbf{Q}_\text{left/right}$ tells us the dimensionality of the subspace of each flavor on the left/right edge of a semi-infinite system (see Fig.~\ref{fig:sweetspot}), and is what we mean by the \textit{chiral charges} in this context. It is furthermore noteworthy that the BBC here is related to the conventional one in two ways: the obvious one is when $p=2$, where no $l>1$ Jordan chain is permitted at the boundary as long as the zero-total-winding restriction is imposed; the other way is to notice that the end/start of a Jordan chain is directly associated with the only right/left eigenvector of the Jordan block, with eigenvalue zero, such that the correspondence between the invariant $\mathbf{W}$ and the numbers of zero modes is actually in this sense maintained.

\section{Generic Boundary Jordan Chains}\label{sec:bjc}

\begin{figure*}[ht]
	\includegraphics[width = 0.95\textwidth]{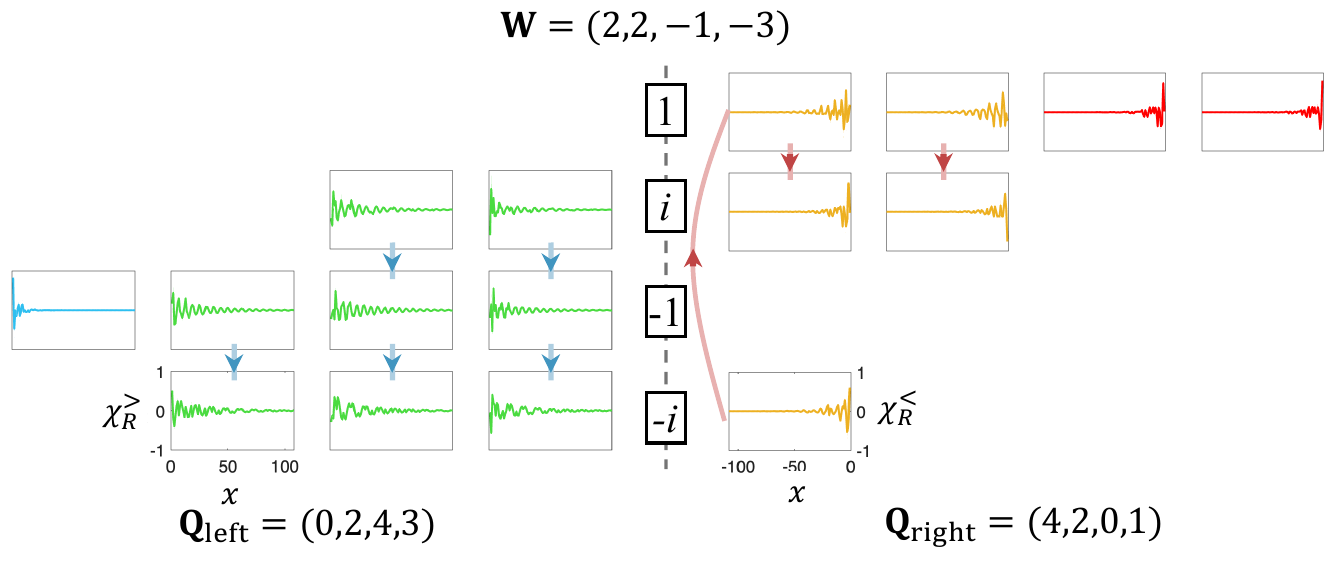}
	\caption{An example of the generic boundary Jordan chains in semi-infinite geometry. Each panel plots a generalized right-eigenvector, $\chi_R^{\gtrless}$ depending on the geometry $x\gtrless 0$, as a function of position $x$ (plotted up to $|x|=108$), obtained and organized according to Eq.~\eqref{eq:jc} (also see App. \ref{sup:bjc}). In the middle of figure, $1, i, -1, -i$ label the chiral flavors (with $p=4$) corresponding to the rows of the panels; on its left the panels plotted in blue and green correspond to the Jordan chains at the left boundary; on its right the panels plotted in red and orange correspond to the Jordan chains at the right boundary. The winding number vector $\mathbf{W}$ is chosen to be the same as in Fig.~\ref{fig:sweetspot}. As a consequence, the panels in this figure, and hence the Jordan chains can be arranged precisely on par with the lower part of Fig.~\ref{fig:sweetspot}. The randomly-generated parameters used in this example are listed in App. \ref{sup:bjc}.}\label{fig:generic_jcs}
\end{figure*}

The sweet-spot model discussed above makes the bulk-boundary correspondence especially transparent. It is still desirable, however, to demonstrate the presence of generic boundary Jordan chains protected by the generalized chiral symmetry when the hopping matrix elements are not fine-tuned as in the sweet-spot model and the loop/chain picture becomes less straightforward in the real space. To that end, we first consider a generic semi-infinite one-dimensional system with $p$-chiral symmetry and compute its boundary Jordan chains as a consequence of its non-trivial winding number vector. We will present the case of a semi-infinite chain terminating on a boundary to the left, and omit the right-boundary case as it is completely analogous.

The problem of finding boundary Jordan chains in a semi-infinite gapped system can be cast into solving a set of generalized Schr\"{o}dinger equations:
\begin{align}\label{eq:seqn}
    (H_{>})^s \psi^{(s)}_R = 0,
    \quad
    (H_{>}^\dag)^s \psi^{(s)}_L = 0,
    \quad (s=1,2,\ldots)
\end{align}
where $H_{>}$ is the Hamiltonian of the semi-infinite system, $\psi^{(s)}_{R/L}$ is the generalized right/left zero mode of rank $s$ (not to be confused with the spatial left and right). A solution of rank $s$ is also a solution of rank $s'>s$. A Jordan chain of length $m$ corresponds to a series of non-vanishing (right- and left-eigenvector) solutions $\{\chi_{R}, (H_{>})\chi_{R}, \ldots, (H_{>})^{m-1}\chi_{R}\}$ and $\{(H_{>}^\dag)^{m-1}\chi_{L},  (H_{>}^\dag)^{m-2}\chi_{L}, \ldots, \chi_{L}\}$ with $(H_{>})^{m}\chi_{R}=(H_{>}^\dag)^{m}\chi_{L}=0$ and $\chi_{L}^\dag (H_{>})^{s-1} \chi_{R}=\delta_{sm}$ $(s=1,\ldots,m)$. Before proceeding, notice that: first, in general $(H_{>})^s \ne (H^s)_{>}$, namely, the $s$-th power of a semi-infinite Hamiltonian is generally not equal to the semi-infinite cutoff of the $s$-th power of an infinite Hamiltonian; second, $H_{>}$ always takes on the form of Eq.~\eqref{eq:pcham} owing to the $p$-chiral symmetry.

The structure of $H_{>}$ as in Eq.~\eqref{eq:pcham} implies that Eq.~\eqref{eq:seqn} can be written in a decoupled form
\begin{align}\label{eq:seqn_j}
    A_{j+s,j}\psi_{Rj}^{(s)} = 0,\quad
    A_{j,j-s}^\dag\psi_{Lj}^{(s)} = 0,
    \quad (s=1,2,\ldots)
\end{align}
where
\begin{align}\label{eq:Adef}
    A_{j+s,j} \equiv \mathbf{a}_{j+s,j+s-1}\cdots\mathbf{a}_{j+1,j},
\end{align}
$j$ labels chiral flavors and $\psi_{R/Lj}^{(s)}$ is the projection of $\psi_{R/L}^{(s)}$ to the $j$-th chiral flavor. Note that each $\mathbf{a}_{j+1,j}$, as well as $A_{j+s,j}$ and $\psi_{R/Lj}^{(s)}$, is semi-infinite. It is clear that a solution of $A_{j+s,j} \psi_{Rj}^{(s)} = 0$ with $s>1$ leads to a solution of $A_{j+s,j+1} \psi_{Rj+1}^{(s-1)} = 0$ with $\psi_{Rj+1}^{(s-1)} = \mathbf{a}_{j+1,j} \psi_{Rj}^{(s)}$, and a solution of $A_{j,j-s}^\dag \psi_{Lj}^{(s)} = 0$ with $s>1$ leads to a solution of $A_{j-1,j-s}^\dag \psi_{Lj-1}^{(s-1)} = 0$ with $\psi_{Lj-1}^{(s-1)} = \mathbf{a}_{j,j-1}^\dag \psi_{Lj}^{(s)}$. Thus a Jordan chain of length $m$ now corresponds to a series of non-vanishing (right- and left-eigenvector) solutions $\{\chi_{Rj}, \chi_{Rj+1}, \ldots, \chi_{Rj+m-1}\}$ and $\{\chi_{Lj}, \chi_{Lj+1}, \ldots, \chi_{Lj+m-1}\}$ that satisfy
\begin{subequations}\label{eq:jc}
\begin{align}
    &A_{j+m,j}\chi_{Rj} = 0, \\
    &\chi_{Rj+\delta+1} = \mathbf{a}_{j+\delta+1,j+\delta} \chi_{Rj+\delta}; \\
    &A_{j+m-1,j-1}^\dag \chi_{Lj+m-1} = 0, \\
    &\chi_{Lj+\delta} = \mathbf{a}_{j+\delta+1,j+\delta}^\dag \chi_{Lj+\delta+1}. \\
    &(\delta = 0,\ldots,m-2) \nonumber
\end{align}
\end{subequations}

\begin{figure*}[ht]
	\centering
	\begin{subfigure}[t]{0.3\textwidth}
		\centering
		\includegraphics[width=\linewidth]{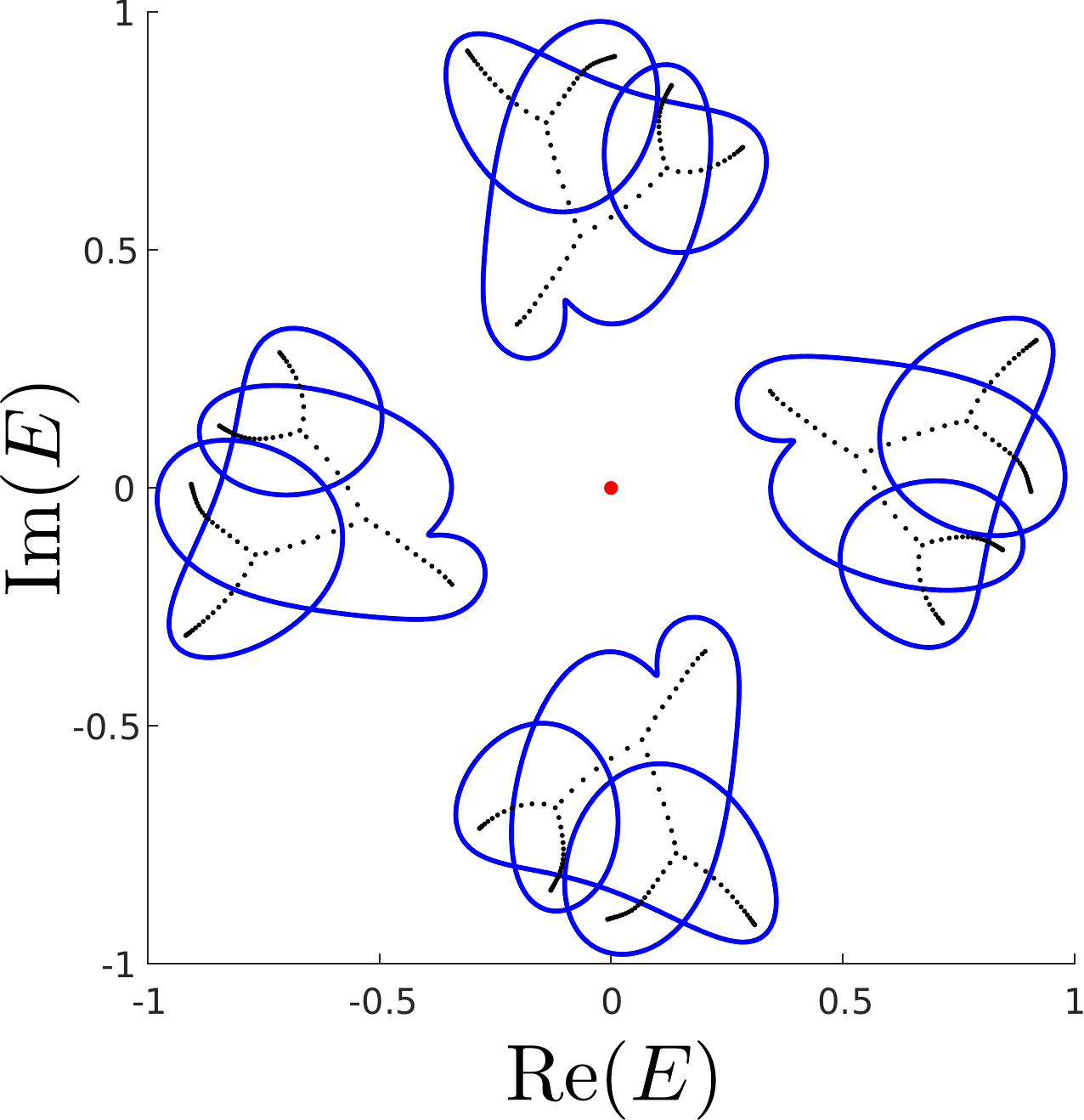}
		\caption{}
	\end{subfigure}
	~
	\begin{subfigure}[t]{0.3\textwidth}
		\centering
		\includegraphics[width=\linewidth]{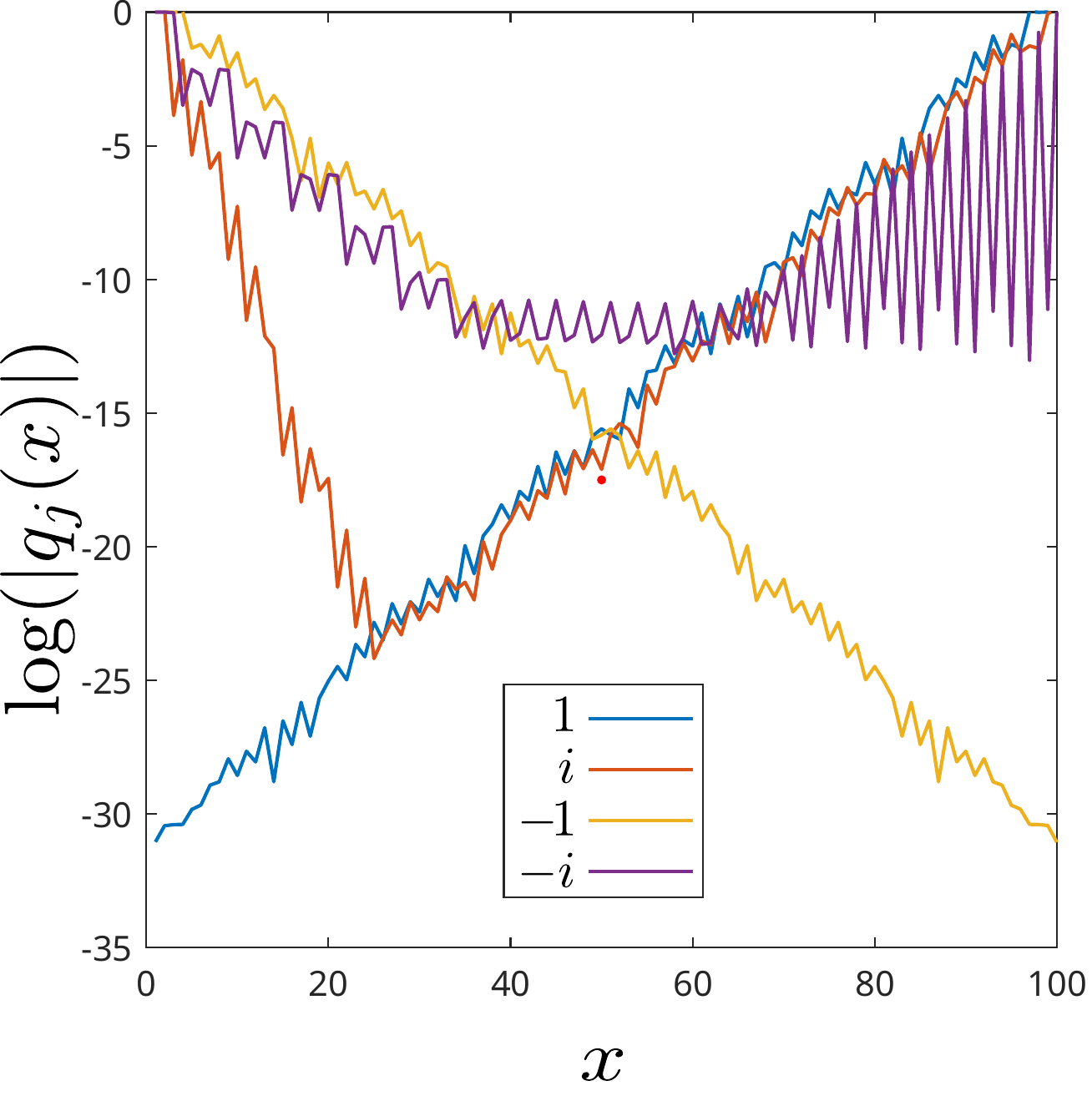}
		\caption{}
	\end{subfigure}
	~
	\begin{subfigure}[t]{0.3\textwidth}
		\centering
		\includegraphics[width=\linewidth]{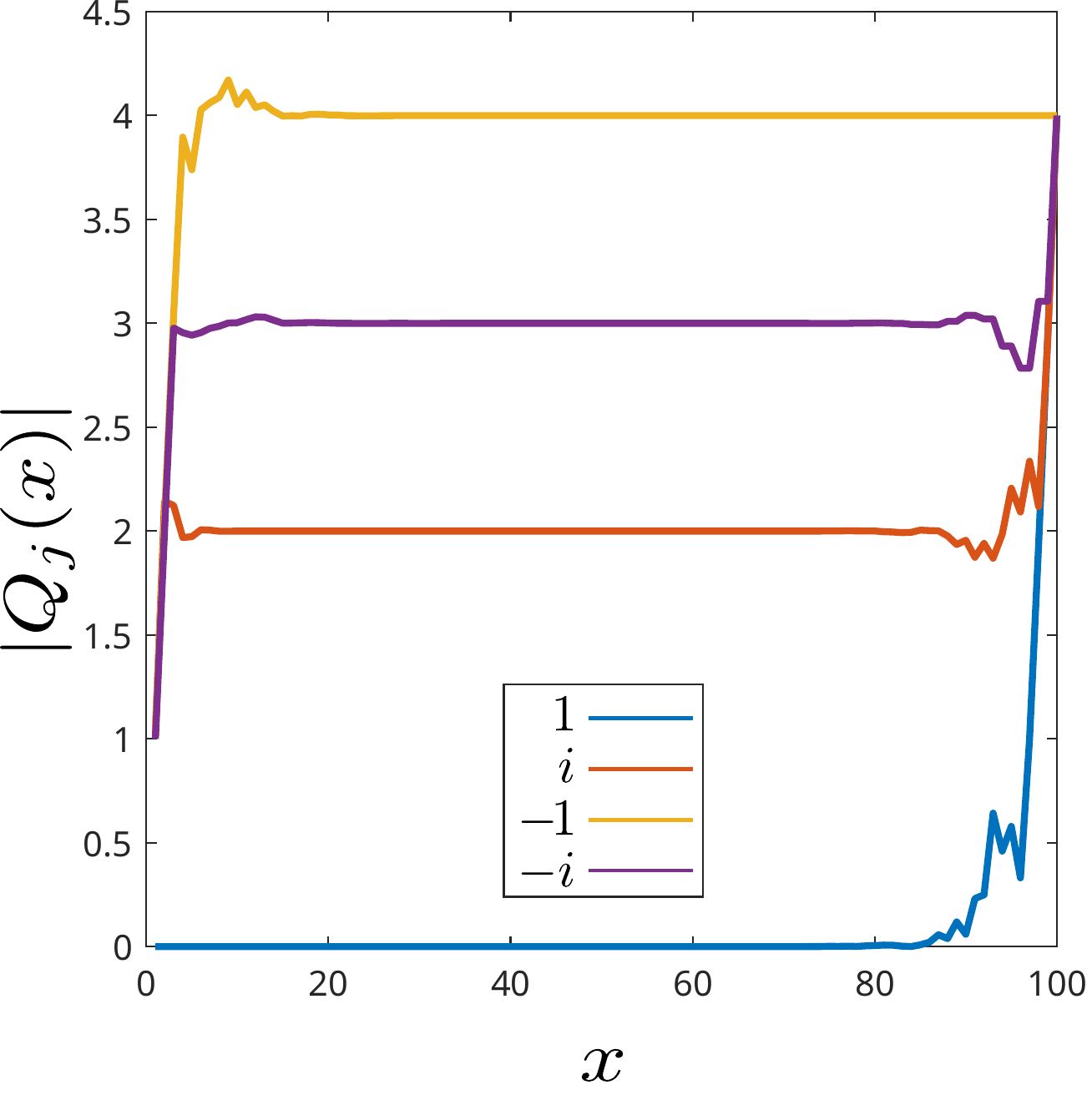}
		\caption{}
	\end{subfigure}
	\caption{(a) The periodic- and open-boundary spectra for a system with $\mathbf{W} = (2, 2, -1, -3)$ plotted with blue lines and black dots, respectively. The zero-energy states are highlighted in red. The model Hamiltonian is randomly constructed as in Fig.~\ref{fig:spectra}. The energies are normalized such that the largest modulus is set to one. The length of the open chain is 100 sites. (b) The logarithm of the modulus of the charge distribution for each chiral flavor, $q_j(x)$ with the label $j$ in $\{1, i, -1, -i\}$, calculated using Eq.~\eqref{eq:charge}. (c) The modulus of the cumulative sums of the charge distribution $Q_j(x) = \sum_{x'=1}^x q_j(x')$. From here, we can see how much charge of each chiral flavor is located at either side of the chain. On the right end, all lines converge to the same value as expected by the theory.}\label{fig:charges}
\end{figure*}

For any fixed pair $j$ and $s$, Eq.~\eqref{eq:seqn_j} can be solved by using the standard transfer matrix technique (see App. \ref{sup:bjc}). As mentioned earlier, the solutions are generally redundant for different $s$ because, for example, a solution $\psi_{Rj}$ satisfying $A_{j+s,j} \psi_{Rj} = 0$ automatically satisfies an equation of a higher rank $s'>s$: $A_{j+s',j} \psi_{Rj} = 0$. Such redundancies can be removed by recursively finding the solutions at each rank $s$ that are orthogonal to the solutions of rank $s-1$ if $s>1$. By organizing all the linearly independent solutions according to Eq.~\eqref{eq:jc}, we can obtain all the boundary Jordan chains in a generic model (see App. \ref{sup:bjc} for details). In Fig.~\ref{fig:generic_jcs}, we show an example with exactly the same winding number vector $\mathbf{W}$ as in Fig.~\ref{fig:sweetspot} but nonvanishing hopping amplitudes up to the third neighbors. We confirm that the generic Jordan chains follow precisely the same boundary-bulk correspondence as discussed in the preceding section.

As soon as we move away from the sweet spot, the boundary modes develop exponentially decaying tails, which can be clearly seen in Fig.~\ref{fig:generic_jcs}. This means that the boundary modes at one end of a finite chain will have a non-vanishing albeit exponentially small coupling with the boundary modes at the other end. In our context, this coupling not only implies the small splitting of exact zero modes as routinely seen in conventional finite-size topological systems, but more importantly leads to a lifting of boundary Jordan blocks to diagonalizable blocks with \textit{almost-zero} eigenvalues. Consequently, the number of the eigenvectors associated with these  almost-zero eigenvalues is given by $\mathbf{Q}_\text{right/left}$ instead of $\mathbf{C}$ ($\mathbf{C}'$) which counts only the ends (starts) of exact Jordan chains. From a mathematical perspective, the nonzero overlap between the boundary modes at each end allow us to match all left and right eigenstates of a given flavor which were previously unbalanced on either end. This splits the defective zero-energy eigenvalue and allows us once again to construct a complete biorthogonal basis diagonalizing the Hamiltonian. To verify this, we numerically compute the chiral charge distribution associated with the almost-zero modes in a generic finite-size chain, defined as
\begin{equation}\label{eq:charge}
	q_j(x) = \text{Tr}\left[V^\dag P_{j,x} U\right],
\end{equation}
where, following the formalism of biorthogonal quantum mechanics \cite{Brody_2013, Edvardsson_2023}, $U$ ($V$) contains all the right (left) eigenvectors with almost-zero eigenvalues, and $P_{j,x}$ is the projector to the $\omega^j_p$ flavor on the $x$-th site. Note that $P_{j,x}$ is not in general biorthogonally Hermitian (see discussion on p.~7 below Eq.~(27) in \cite{Brody_2013}), and therefore $q_j(x)$ is not necessarily real; however, our calculations find that the imaginary part of a sum of $q_j(x)$ starting from a boundary site is exponentially suppressed as we include more and more sites in the sum. In Fig.~\ref{fig:charges}, we show one example constructed in the same way as the models used for Fig.~\ref{fig:spectra}, that has the same $\mathbf{W}$ as the above sweet-spot example. The gapped PBC and OBC spectra in this non-sweet-spot model are shown in Fig.~\ref{fig:charges}(a). The modulus of the spatial charge distribution, $|q_j(x)|$, is plotted in Fig.~\ref{fig:charges}(b) on a logarithmic scale for each chiral flavor, which clearly indicates that the almost-zero modes are exponentially localized at the edges. In Fig.~\ref{fig:charges}(c), we show the moduli of the cumulative sums $Q_j(x) = \sum_{x'=1}^x q_j(x')$ for each flavor. This provides a clearer picture of how many charges of each flavor is localized to which edge of the chain, and is fully consistent with the counting procedure as derived in App.~\ref{sup:algo}. In App.~\ref{sup:charge}, we further show how the boundary charge is preserved in time despite not only containing eigenvectors.

Three extra remarks are in order: first, given that $p=4$ in the above example is a composite number, we can break the symmetry down to a $p=2$ symmetry as we mentioned in the final paragraph of Sec.~\ref{sec:windtop}. In this case, the initial chiral flavors combine into the new flavors according to $(1, -1)\to 1$ and  $(i, -i)\to -1$. The stable boundary charges would then simply be the remaining charge imbalance on either edge. Using initial $\mathbf{W} = (2, 2, -1, -3)$ again for example, we would have $(0+4, 2+3) \rightarrow (0, 1) = \mathbf{Q}_\text{left}^\text{(new)}$ and $(4+0, 2+1) \rightarrow (1, 0) = \mathbf{Q}_\text{right}^\text{(new)}$, which is obviously consistent with $\mathbf{W}^\text{(new)} = (1, -1)$. For a more general discussion on this, see App.~\ref{sup:composite}. Second, note that the same set of winding numbers arranged in a different order will yield a completely different charge distribution at the edges; for example, $\mathbf{W} = (2,-1,2,-3)$ would only have three charges in total for each chiral flavor.

\section{Summary and Discussion}
In this work, we have studied the topology of systems with a generalized chiral symmetry. The symmetry is an extension of the Bernard-LeClair symmetry classification and can be seen as  an additional axis for the periodic table of topological phases labeled by integers $p$. A notable new feature is that the topological phase is specified by a vector of winding numbers rather than a single integer. Meanwhile, the BBC in one dimension is promoted to a correspondence between the vector of winding numbers and the set of Jordan chains that underlie the boundary Hilbert space. This implies, on the one hand, that the boundary chiral charges can exhibit highly unconventional patterns: the boundary charges at the two edges are complementary but not necessarily equal whereas the maximum number at either edge can well exceed the largest magnitude of the winding numbers. On the other hand, the non-diagonalizability associated with boundary Jordan chains of length larger than $1$ is topologically protected in a semi-infinite setting, suggesting that it is not sufficient to look at only eigenvectors when studying topological modes, but rather the zero-energy subspace which also contains generalized eigenvectors of rank higher than one.

The obvious next question theory-wise is what happens in higher dimensions; as seen in App.~\ref{sup:wind}, the relation between the overall winding number and the winding number of the subblocks holds true in all odd dimensions, which is highly suggestive of there being related $p$CS phases lurking there. However, the presence/absence of skin modes is no longer merely determined by the overall winding number, but also by weak invariants \cite{Okuma_2020, Zhang_2020, Zhang_2022}, so extra care needs to be taken. It is also expected that the BBC must be revised, and the nature of the topological modes would most likely no longer be simple Dirac-type cones.

For $p$CS to gain relevance in experiments, we propose implementations based on current technology in acoustic \cite{Zhang_2021} or photonic systems \cite{Longhi_2015a, Longhi_2015b, Longhi_2018, Viedma_2023, Roccati_2024}. Indeed, a gapless version of a $p$CS chain has already been realized by using photonic rings \cite{Viedma_2023}. Another convenient platform to simulate the relevant physics is to construct an electric circuits with a Laplacian that mimics the non-Hermitian Hamiltonian through the appropriate combination of electric components \cite{Hofmann_2019, Zhang2_2022, Zeng_2022, Liu_2021, Zou_2021}. It will be intriguing to experimentally detect the boundary Jordan chains implied by the $p$CS and its associated BBC discovered in this paper.

\acknowledgments
We are grateful to Wenbin Rui for very helpful discussion. This work was supported by the National Natural Science Foundation of China under Grants No. 92265201 and No. 12574176, and the Innovation Program for Quantum Science and Technology under Project No. 2021ZD0302704.

\appendix

\section{Winding Numbers}\label{sup:wind}

Even though we are only considering one-dimensional systems in this work, we show here that the total winding number is always a sum of the individual blocks in all odd dimensions. The $(2n+1)$-dimensional spectral winding number around zero is given by  \cite{Budich_2013, Schnyder_2008}
\begin{equation}
	\begin{split}
		&w_{2n+1}(H_{\mathbf{k}}) = \frac{(-1)^n n!}{(2n+1)!}\left(\frac{i}{2\pi}\right)^{n+1} \varepsilon^{\alpha_1\ldots\alpha_{2n+1}}\times\\
		&\int_{BZ}\text{Tr}\left[H_{\mathbf{k}}^{-1}\left(\partial_{\alpha_1}H_{\mathbf{k}}\right)\ldots H_{\mathbf{k}}^{-1}\left(\partial_{\alpha_{2n+1}}H_{\mathbf{k}}\right)\right]d^{2n+1}k.
	\end{split}
\end{equation}
We observe that the above formula only contains products of the form $H^{-1}\partial H$. Calculating such a product for the Hamiltonian in Eq.~\eqref{eq:pcham}, gives us
\begin{equation}
	\begin{split}
		&H_{\mathbf{k}}^{-1}\partial_\alpha H_{\mathbf{k}} = \\
		&\text{diag}\left[\mathbf{a}_{2,1}^{-1}\partial_\alpha \mathbf{a}_{2,1}, \ldots, \mathbf{a}_{p,p-1}^{-1}\partial_\alpha \mathbf{a}_{p,p-1}, \mathbf{a}_{1,p}^{-1}\partial_\alpha \mathbf{a}_{1,p}\right],
	\end{split}
\end{equation}
which means that any products of such terms will remain block diagonal, and when we finally perform the trace, it can be achieved by summing the traces of all the individual blocks together.

\section{Constructing Non-Island Toy Models}\label{sup:toy}

\begin{figure}
	\includegraphics[width=0.9\linewidth]{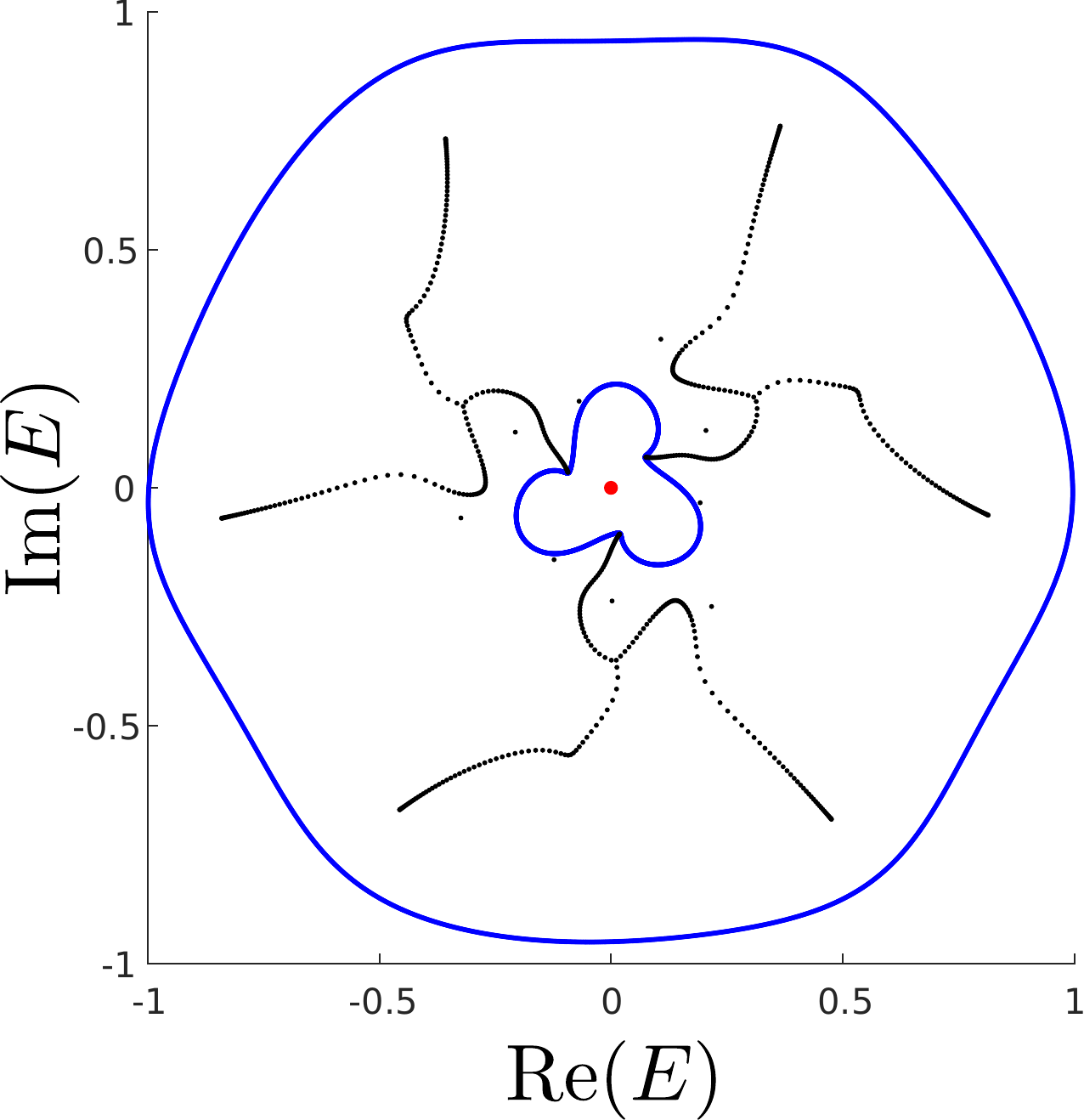}
	\caption{A randomly generated example of a 3-CS system with a periodic-boundary spectrum not given by isolated islands. The blue lines are the periodic-boundary spectrum, the black points are the open-boundary energies, with the exception that the zero-energy eigenvalue is colored in red. The number of sites in the open system is 100. The energies have been rescaled such that the largest energy modulus is set to one.}\label{supfig:nonisland}
\end{figure}

The toy model used in the main text can only give band islands, meaning the spectra will separate into individual bands. This is more directly connected to the conventional Hermitian AIII case, since it consists of two band islands lying on the real line that are mapped to each other via the chiral operator. However, for non-Hermitian systems, this island picture is not a necessary feature, and we can, in fact, also conjure up models where we, for example, have two bands circling the origin but in opposite directions such that the overall winding remains zero. This can be achieved by starting from a block
\begin{equation}
	\mathbf{a}_0(k) =
	\begin{pmatrix}
		a_{11} e^{ik} & a_{12}\\
		a_{21} & a_{22}e^{-ik}
	\end{pmatrix}.
\end{equation}
When the off-diagonal terms are zero, the matrix decouples into two blocks with opposite winding.  
We can then construct the remaining blocks using for example the template
\begin{equation}
	\mathbf{a}_{mn}(k) =
	\begin{pmatrix}
		A_{11} e^{imk} & A_{12}\\
		A_{21} & A_{22} e^{ink}
	\end{pmatrix},
\end{equation}
where the winding is $m+n$ as long as $\vert A_{12}A_{21}\vert < \vert A_{11}A_{22}\vert$. This can be confirmed by a direct calculation of the winding number using Eq.~\eqref{eq:winding}. We provide an illustrative example in Fig.~\ref{supfig:nonisland}, where we have a randomly generated 3-CS model with $\mathbf{W} = (1,-1,0)$ (with $(m_1,n_1) = (2,-1)$ and $(m_2,n_2) = (-2,1)$ for the respective two blocks that are not $\mathbf{a}_0$) where the spectrum wraps around the origin, and the OBC modes are confined to be within the annulus with the two PBC bands as its borders. The sampling is done using the following steps: first we find $\vert A_{12}A_{21}\vert$ and $\vert A_{11}A_{22}\vert$ in the same way as outlined for $\vert c\vert$ and $\vert d\vert$ in the main text. Then we define $\vert A_{12}\vert = r\vert A_{12}A_{21}\vert$ and $\vert A_{21}\vert = \vert A_{12}A_{21}\vert /r$ using a uniformly sampled $r$ from $[0.5,2]$. Then we sample the phases from $[-\pi, \pi]$ as before. An identical procedure is performed on the diagonal terms. As expected, we still have topological zero modes at the edges. 

\section{Bulk-Boundary Correspondence}\label{sup:charge}

Here we will show that the chiral charge Eq.~\eqref{eq:charge} is preserved at the edge of a semi-infinite chain. In the semi-infinite case, the Hamiltonian is not necessarily diagonalizable, so the zero-energy subspace consists not only of eigenstates, but also of higher-order eigenstates (i.e. eigenstates of powers of the Hamiltonian). To calculate the $n$th chiral charge at time $t$, we must then look at the expression
\begin{equation}
	\begin{split}
		Q(n, t) &= \text{Tr}\left[V(t)^\dag \mathcal P_n U(t)\right]\\
		&= \text{Tr}\left[V^\dag e^{iHt}\mathcal P_n e^{-iHt}U\right],
	\end{split}
\end{equation}
where 
\begin{equation}\label{supeq:proj}
	\mathcal{P}_n = \frac{\prod_{q\neq n}\left(\omega^q_p - C_p\right)}{\prod_{r\neq n}\left(\omega^r_p - \omega^n_p\right)}
\end{equation}
is the projector to the $n$th chiral-charge sector, and $U$ and $V$ are matrices containing the right and left (including higher order) zero-energy  eigenvectors, respectively. Note that they are all eigenvectors of the chiral operator.

We insert $\mathbb I = U(t)V(t)^\dag$ between all the terms in the product of Eq.~\eqref{supeq:proj}. For each term, we get
\begin{equation}
	\begin{split}
		V(t)^\dag \left(\omega^q_p - C_p\right)U(t) &= \omega^q_p - V^\dag e^{iHt}C_pe^{-iHt}U\\
		&= \omega^q_p - V^\dag e^{iHt}e^{-i\omega_p Ht}C_p U\\
		&= \omega^q_p - V^\dag e^{iHt(1-\omega_p)}U D\\
		&= \omega^q_p - e^{iJt(1-\omega_p)} D,
	\end{split}
\end{equation}
where $D$ is a diagonal matrix containing the eigenvalues of the vectors in $U$ with respect to the chiral operator ($C_p U = UD$). We have denoted the Jordan normal form of the zero-energy subspace of the Hamiltonian by $J$. This matrix can be further decomposed into one zero-matrix $\mathbf{0}$ corresponding to all the one-dimensional Jordan blocks, and another one $J_1$ containing the Jordan blocks of higher dimension. Since $J = \mathbf 0\oplus J_1$, we can treat them independently. For $\mathbf{0}$, the exponent reduces to an identity matrix and we straightforwardly obtain the number of chiral charges with flavor $n$ in that sector of the zero-energy subspace. For $J_1$, the story is essentially the same, but we must further note that  $e^{iJ_1t(1-\omega_p)}$ contains a finite number of terms in its series expansion, but only the zeroth order term -- the identity matrix -- lies on the diagonal. The rest of the nonzero terms lie above the diagonal. This means that once we multiply all the terms together and take the trace, only the zeroth-order terms survive, so the conclusions are the same as that of $\mathbf{0}$. Putting it all together, we get
\begin{equation}
	Q(n, t) = Q(n, 0).
\end{equation}

This means that although higher-order eigenvectors will move away from being chiral eigenvectors, the total charge of any particular flavor is preserved. This may seem paradoxical, but it can be understood by remembering that the higher-order left and right eigenvectors move in ``opposite'' directions: one evolves into higher chiral flavors corresponding to higher powers of $\omega_p$, while the other evolves into lower flavors with their only overlap lying in the original flavor sector.

\section{Deriving the Chiral Charge Distribution}\label{sup:algo}

To determine the charge of each flavor on a given boundary of the 1-D system, we first focus on the sweet-spot case as explained in the main text and then consider the more general case. To make the discussion clearer, we use the sublattice picture for the different flavors and adjust our language accordingly. 

\subsection{The Sweet-Spot Model}

In the sweet-spot model, the winding number $\mathcal{W}(\mathbf{a}_{j+1,j})$ associated with the $j+1$ and $j$ sublattices determines the relative position of the site in the $(j+1)$th sublattice with respect to the position of a site in the $j$th sublattice to which the former site is connected through the hopping term in the Hamiltonian. For example, site $x$ of the $j$th sublattice connects to the site $x+2$ in the $(j+1)$th sublattice if $\mathcal{W}(\mathbf{a}_{j+1,j}) = 2$. Because the existence of a point gap at zero energy requires the sum of all winding numbers to be zero, and each winding number equals the relative position translation, all such hoppings between sublattices form closed loops in the bulk. The presence of a boundary then cuts some of these loops, leaving behind isolated sites and open chains. These contribute to the boundary charges. It is worthwhile to keep Fig.~\ref{fig:sweetspot} from the main text in mind throughout this discussion.

All loops in the bulk have the same shape as determined by the vector invariant in Eq.~\eqref{eq:vectorinv}. Consequently, the number of sites not belonging to a complete loop on the left boundary of a given sublattice equals the number of sites to the left of the first site  in that sublattice that is a member of a complete loop. In other words, the charge number on the left boundary of that flavor is determined by the relative position of a site in that sublattice with respect to the left-most site of the closest complete loop. For the right edge, we can make analogous statements.

To explicitly determine the left-side boundary charges for a given $\mathbf{W}$, we begin by denoting the position of the first flavor in the left-most complete loop by $x_1$. The positions of the remaining $p-1$ flavors in the loop are then given by 
\begin{equation}
	x_n = x_1 + \sum_{j=1}^{n-1} W_j \equiv x_1 + (\mathbf{Q})_n,
\end{equation}
where $\mathbf{Q}$ is defined as a vector of the cumulative sums of the elements of $\mathbf{W}$ starting with $(\mathbf{Q})_1 = 0$ (or, $\mathbf{Q}$ can be viewed alternatively as the vector of the cumulative sums cyclically permuted so that $(\mathbf{Q})_1 = \sum_{j=1}^p W_j = 0$). The left-most position of the loop is then 
\begin{equation}
	\min_n \left\{x_n\right\} = x_1 + \min \mathbf{Q}.
\end{equation}
The left-most position of the left-most loop must coincide with the left boundary which is defined to be $0$, so $x_1 + \min \mathbf{Q} = 0$, or $x_1 = -\min \mathbf{Q} \ge 0$. Then the positions -- and hence the total left-boundary charge -- of a given flavor is simply given by
\begin{equation}
	\mathbf{Q}_\text{left} = x_1 + \mathbf{Q} = -\min \mathbf{Q} + \mathbf{Q}.
\end{equation}
For the right charges, we can do a similar analysis, using the maximum instead of the minimum. 

\subsection{The General Case}

In a general model with semi-infinite geometry (say, one with a left boundary), let us assume such a Hamiltonian, $H_{>}$, can be brought to its Jordan canonical form by a similarity transformation. The winding number vector $\mathbf{W}$ determines the number of boundary right/left eigenvectors (with eigenvalue $0$) of each flavor as in the usual Hermitian case, but the problem is to find the remaining dimensions associated with the defective Jordan blocks (with eigenvalue $0$).

As explained in the main text, the generalized eigenvectors of eigenvalue zero can always be chosen to be an eigenvector of the chiral operator, namely, they each belong to a specific chiral flavor. Let us denote the total number of ordinary left and right eigenvectors of eigenvalue zero for chiral flavor $j$ by $l_j$ and $r_j$, respectively. In a similar fashion, denote the number of generalized eigenvectors associated with the zero-energy subspace and flavor $j$ by $\delta l_j$ and $\delta r_j$. Further note that a conventional zero mode contributes as both a left and a right eigenvector. Using the conventional bulk-boundary correspondence, we can establish that
\begin{equation}\label{eq:lr_app}
    \begin{split}
    l_j &= \max(0, W_{j-1}),\\
    r_j &= \max(0, -W_{j}) = -\min(0, W_{j}).
    \end{split}
\end{equation}
On the other hand, the total dimension of the boundary Hilbert space for each chiral flavor, i.e. the total number of generalized eigenvectors of eigenvalue zero for each chiral flavor, $c_j$, is identical by counting either the left generalized eigenvectors or the right ones. Formally, we write
\begin{equation}\label{eq:cj_app}
    c_j = l_j + \delta l_j = r_j + \delta r_j.
\end{equation}
Now the crucial step is to notice that any left generalized eigenvector $\psi_{L,j}$, that is counted in $\delta l_j$ (meaning it can generate another left generalized eigenvector $\psi_{L,j-1}$ by the action of $H_{>}^\dag$), has its dual $\psi_{R,j}$ belonging to $\delta r_{j-1}$, because $\psi_{R,j}$ is generated by the action of $H_{>}$ on $\psi_{R,j-1}$, which is the dual of $\psi_{L,j-1}$. This map is invertible, yielding the equation
\begin{equation}\label{eq:dldr_app}
    \delta l_j = \delta r_{j-1}.
\end{equation}
This equation, via Eq.~\eqref{eq:cj_app}, is consistent with the simple fact $\sum_j l_j = \sum_j r_j$, which counts the total number of defective Jordan blocks including ordinary zero modes. Furthermore, Eqs.~(\ref{eq:lr_app}-\ref{eq:dldr_app}) lead to a recursive relation
\begin{equation}\label{eq:rec_app}
     c_{j+1} = c_j + (l_{j+1} - r_j) = c_j + W_j,
\end{equation}
where we have used $\max(0, W_{j}) + \min(0, W_{j}) = W_j$. This implies
\begin{equation}
     c_{j} = c_1 + (\mathbf{Q})_j,\qquad
     (\mathbf{Q})_j \equiv \sum_{j'=1}^{j-1} W_{j'},
\end{equation}
which leaves a single variable (in our case $c_1$) to be fixed to completely determine $\mathbf{Q}_\text{left} = (c_1, c_2, \ldots, c_p)$.

To fix the only remaining free variable, we can invoke the topological stability argument: if all $c_j$s are nonzero (i.e. a positive integer), then we can gap out boundary modes by adding local perturbation such that all $c_j$'s decrease by the same amount (because a generic eigenvector of nonzero eigenvalue in this system must involve all chiral flavors) until we are left with at least one flavor $j_0$ with $c_{j_0}=0$ (cf. the first flavor in the example illustrated in Fig.~\ref{fig:sweetspot}). This argument also implies that $(\mathbf{Q})_{j_0} = \min \mathbf{Q}$ and thus $c_{j_0} = c_1 + \min \mathbf{Q} = 0$. It then follows immediately that
\begin{equation}
     c_{j} = -\min \mathbf{Q} + (\mathbf{Q})_j.
\end{equation}
That is, $\mathbf{Q}_\text{left} = -\min \mathbf{Q} + \mathbf{Q}$. A similar heuristic proof can be formulated for a general model with a semi-infinite chain terminating on a boundary to the right to show that $\mathbf{Q}_\text{right} = \max \mathbf{Q} - \mathbf{Q}$.

\section{Solutions of Boundary Jordan Chains in Generic Semi-infinite Systems}\label{sup:bjc}

We start from a generic Hamiltonian satisfying $p$CS,
\begin{align}
	H = \sum_{j=1}^{p}\sum_{i}\sum_{n=-n_c}^{n_c} \ket{j+1,i+n}a_{j+1,j;i+n,i}\bra{j,i},
\end{align}
where $i$ labels the sites in the one-dimensional system, $n$ labels its neighbors, $n_c$ is the cutoff range of neighbors, and $a_{j+1,j;i+n,i}$ is the hopping amplitude from site $i$ in the $j$th chiral flavor to site $i+n$ in the $(j+1)$th chiral flavor. We omit the inclusion of any inner degrees of freedom as the following analysis would remain largely the same, albeit with more indices cluttering up the exposition. With the translational invariance, $a_{j+1,j;i+n,i} \equiv a_{j+1,j;n}$ does not depend on $i$. But for a semi-infinite system, the Hilbert space is spanned by $\{\ket{j,i>0}\}$ ($\{\ket{j,i<0}\}$) if the left (right) boundary is to be considered. In all cases, the Hamiltonian preserves the form of Eq.~\eqref{eq:pcham}.

As long as $n_c$ is finite, we can always enlarge the unit-cell to contain $n_c$ sites such that the effective hopping range becomes $1$, thus for a semi-infinite geometry ($i\ge 1$), the blocks in Eq.~\eqref{eq:pcham} can be expressed as
\begin{equation}\label{eq:mataj}
	\mathbf{a}_{j+1,j} =
	\begin{pmatrix}
		\mathbf{a}_{j+1,j;0} & \mathbf{a}_{j+1,j;-1} & 0 & 0 & 0 \\
		\mathbf{a}_{j+1,j;1} & \mathbf{a}_{j+1,j;0} & \mathbf{a}_{j+1,j;-1} & 0 & 0 \\
		0 & \mathbf{a}_{j+1,j;1} & \mathbf{a}_{j+1,j;0} & \mathbf{a}_{j+1,j;-1} & 0 \\
		0 & 0 & \ddots & \ddots & \ddots
	\end{pmatrix},
\end{equation}
where
\begin{align}
	&\mathbf{a}_{j+1,j;0} = 
	\begin{pmatrix}
		a_{j+1,j;0} & a_{j+1,j;-1} & \cdots & a_{j+1,j;1-n_c} \\
		a_{j+1,j;1} & a_{j+1,j;0} & \ddots & a_{j+1,j;2-n_c} \\
		\vdots & \ddots & \ddots & \ddots \\
		a_{j+1,j;n_c-1} & \ddots & a_{j+1,j;1} & a_{j+1,j;0}
	\end{pmatrix}, \\
	&\mathbf{a}_{j+1,j;1} = 
	\begin{pmatrix}
		a_{j+1,j;n_c} & a_{j+1,j;n_c-1} & \cdots & a_{j+1,j;1} \\
		0 & a_{j+1,j;n_c} & \ddots & a_{j+1,j;2} \\
		0 & 0 & \ddots & \ddots \\
		0 & 0 & 0 & a_{j+1,j;n_c}
	\end{pmatrix}, \\
	&\mathbf{a}_{j+1,j;-1} = 
	\begin{pmatrix}
		a_{j+1,j;-n_c} & 0 & 0 & 0 \\
		a_{j+1,j;1-n_c} & a_{j+1,j;-n_c} & 0 & 0 \\
		\vdots & \ddots & \ddots & 0 \\
		a_{j+1,j;-1} & a_{j+1,j;-2} & \ddots & a_{j+1,j;-n_c}
	\end{pmatrix}.
\end{align}

To solve the equation
\begin{align}\label{eq:apsij}
	\mathbf{a}_{j+1,j}\psi_{Rj} = 0,
\end{align}
which is a special case of Eq.~\eqref{eq:seqn_j} for the right eigenvector with $s=1$, we first write it explicitly by using Eq.~\eqref{eq:mataj} as
\begin{align}
	&\mathbf{a}_{j+1,j;1}\psi_{Rj;m-1} + \mathbf{a}_{j+1,j;0}\psi_{Rj;m} + \mathbf{a}_{j+1,j;-1}\psi_{Rj;m+1} \nonumber\\
	&\quad = 0, \qquad (m>1) \label{eq:tm0}\\
	&\mathbf{a}_{j+1,j;0}\psi_{Rj;1} + \mathbf{a}_{j+1,j;-1}\psi_{Rj;2} = 0, \label{eq:tm1}
\end{align}
where $\psi_{Rj;m}$ ($m\ge 1$) is a subvector of $\psi_{Rj}$. We can now cast Eq.~\eqref{eq:tm0} into the standard transfer-matrix form
\begin{align}
	\begin{pmatrix}
		\psi_{Rj;m} \\
		\psi_{Rj;m+1}
	\end{pmatrix} = T
	\begin{pmatrix}
		\psi_{Rj;m-1} \\
		\psi_{Rj;m}
	\end{pmatrix},\quad (m>1) \\
	T \equiv
	\begin{pmatrix}
		0 & \mathds{1} \\
		-\mathbf{a}_{j+1,j;-1}^{-1}\mathbf{a}_{j+1,j;1} & -\mathbf{a}_{j+1,j;-1}^{-1}\mathbf{a}_{j+1,j;0}
	\end{pmatrix}.
\end{align}
Note that $\mathbf{a}_{j+1,j;-1}$ need not always be invertible, but we can always add an arbitrarily small perturbation to ensure its invertibility. For in-gap solutions, the transfer matrix $T$ can be generically decomposed into a sum of two parts, $T = T_{|\lambda|<1} + T_{|\lambda|>1}$, where
\begin{align}
	&T_{|\lambda|<1} \equiv
	\begin{pmatrix}
		v_1 & v_2 & \cdots
	\end{pmatrix}
	\begin{pmatrix}
		\lambda_1 &  & \\
		& \lambda_2 & \\
		&  & \ddots
	\end{pmatrix}
	\begin{pmatrix}
		u_1^\dag \\ u_2^\dag \\ \vdots
	\end{pmatrix}, \\
	&T_{|\lambda|>1} \equiv
	\begin{pmatrix}
		v_1^\prime & v_2^\prime & \cdots
	\end{pmatrix}
	\begin{pmatrix}
		\lambda_1^\prime &  & \\
		& \lambda_2^\prime & \\
		&  & \ddots
	\end{pmatrix}
	\begin{pmatrix}
		{u_1^\prime}^\dag \\ {u_2^\prime}^\dag \\ \vdots
	\end{pmatrix},
\end{align}
with $\{v_1, v_2, \ldots\}$ being the eigenvectors of $T$ with eigenvalues $\{\lambda_1, \lambda_2, \ldots\}$ that are less than one in modulus, and $\{v_1^\prime, v_2^\prime, \ldots\}$ being the eigenvectors of $T$ with eigenvalues $\{\lambda_1^\prime, \lambda_2^\prime, \ldots\}$ that are larger than one in modulus. The vectors $u_\alpha$ and $u_\alpha^\prime$ are the corresponding left eigenvectors biorthogonal to the right eigenvectors $v_\alpha$ and $v_\alpha^\prime$, i.e., they satisfy $u_\alpha^\dag v_\beta = {u_\alpha^\prime}^\dag {v_\beta^\prime} = \delta_{\alpha\beta}$, and $u_\alpha^\dag {v_\beta^\prime} = {u_\alpha^\prime}^\dag {v_\beta} = 0$. Any solution $\nu$ to Eq.~\eqref{eq:apsij} in a semi-infinite geometry (with a left end) is then a linear combination of $\{v_1, v_2, \ldots\}$, that also satisfies the boundary condition in Eq.~\eqref{eq:tm1}, i.e.
\begin{align}
	\begin{pmatrix}
		\mathbf{a}_{j+1,j;0} & \mathbf{a}_{j+1,j;-1}
	\end{pmatrix}
	\begin{pmatrix}
		v_1 & v_2 & \cdots
	\end{pmatrix}
	\nu_r = 0,
	\quad (r=1,\ldots,N_j),
\end{align}
where $N_j$ denotes the number of solutions. $N_j$ equals the number of vectors $v_\alpha$ minus the number of constraints imposed by the boundary condition. Since $\mathbf{a}_{j+1,j;n}$ is an $n_c\times n_c$ matrix, the boundary condition consists of $n_c$ equations. The number of $v_\alpha$ is equal to the number $N_{\vert \lambda\vert < 1}$ of eigenvalues of $T$ that lie within the unit circle. This number can be calculated by integrating the function $f(z) = (2\pi i)^{-1}\sum_\lambda (z-\lambda)^{-1}$ along a contour that goes around the unit circle. Here, the sum is over all eigenvalues $\lambda$ of $T$. We can rewrite the function in the form $f(z) = (2\pi i)^{-1}\det(z-T)^{-1}\partial_z \det(z-T)$, from which it follows that
\begin{widetext}
\begin{align}
    N_{\vert \lambda\vert < 1} &= \frac{1}{2\pi i}\oint  \frac{\partial_z \det(z-T)}{\det(z-T)}dz \nonumber\\
    &= \frac{1}{2\pi i}\oint  \text{Tr}\left[\left(z(\mathbf{a}_{j+1,j;0} + z\mathbf{a}_{j+1,j;-1}) + \mathbf{a}_{j+1,j;1}\right)^{-1}\partial_z \left(z(\mathbf{a}_{j+1,j;0} + z\mathbf{a}_{j+1,j;-1}) + \mathbf{a}_{j+1,j;1}\right)\right]dz \nonumber\\
    &= \frac{1}{2\pi i}\oint \frac{\text{Tr}[\mathds{1}] dz}{z} +\frac{1}{2\pi i}\oint  \text{Tr}\left[\left(\mathbf{a}_{j+1,j;0} + z\mathbf{a}_{j+1,j;-1} + \frac{\mathbf{a}_{j+1,j;1}}{z}\right)^{-1} \partial_z \left(\mathbf{a}_{j+1,j;0} + z\mathbf{a}_{j+1,j;-1} + \frac{\mathbf{a}_{j+1,j;1}}{z}\right)\right]dz \nonumber\\
    &= n_c +\frac{1}{2\pi i}\int_{BZ}  \text{Tr}\left[\left(\mathbf{a}_{j+1,j;0} + e^{ik}\mathbf{a}_{j+1,j;-1} + e^{-ik}\mathbf{a}_{j+1,j;1}\right)^{-1}\partial_k \left(\mathbf{a}_{j+1,j;0} + e^{ik}\mathbf{a}_{j+1,j;-1} + e^{-ik}\mathbf{a}_{j+1,j;1}\right)\right]dk \nonumber\\
    &\equiv n_c +\frac{1}{2\pi i}\int_{BZ}  \text{Tr}\left[\mathbf{a}(k)^{-1}_{j+1,j}\partial_k \mathbf{a}_{j+1,j}(k)\right]dk \nonumber\\
    &= n_c + W_j,
\end{align}
\end{widetext}
where $\mathbf{a}_{j+1,j}(k)$ is the corresponding Fourier term in a periodic chain consisting of the enlarged unit cells.

We can now write $N_j = N_{\vert \lambda\vert < 1} - n_c = W_j$. From this we can see that if $W_j < 0$, the equations become overconstrained, and no solutions exist in general, which is consistent with the idea that a negative winding number means that the eigenstates reside on the opposite end of the chain. Similarly, a positive winding number corresponds to states on this end of the chain, with the number of possible solutions equaling said winding number. The $N_j$ solutions to Eq.~\eqref{eq:apsij} are given by
\begin{align}
	\begin{pmatrix}
		\psi_{Rj;m}^{(s=1;r)} \\
		\psi_{Rj;m+1}^{(s=1;r)}
	\end{pmatrix} =
	\begin{pmatrix}
		v_1 \lambda_1^{m-1} & v_2 \lambda_2^{m-1} & \cdots
	\end{pmatrix}
	\nu_r \nonumber\\
	(m\ge 1;\; r=1,\ldots,N_j).
\end{align}

The same strategy can be applied to solve Eq.~\eqref{eq:seqn_j} for all $j$ and $s$. The remaining issues are to remove the redundancy in the solutions and to arrange the solutions into Jordan chains. The former can be achieved by finding the complement of $\mathrm{Span}(\{\psi_{R/Lj}^{(s;r)}\})$ in $\mathrm{Span}(\{\psi_{R/Lj}^{(s+1;r)}\})$ for each $j$ and $s$, namely, finding linear combinations of $\{\psi_{R/Lj}^{(s+1;r)}\}$ that lie completely outside the span of $\{\psi_{R/Lj}^{(s;r)}\}$. In general, finding the complement of $\mathrm{Span}(B)$ in $\mathrm{Span}(A)$, where $A$ and $B$ are both sets consisting of linearly independent vectors, and $\mathrm{Span}(B)\subseteq \mathrm{Span}(A)$, can be done as follows: first solve the linear equation $\mathbf{A}X=\mathbf{B}$, where $\mathbf{A}$ and $\mathbf{B}$ are matrices, whose columns are the vectors in $A$ and $B$, respectively; then a nonzero vector $\mathbf{A}y \in \mathrm{Span}(A)$ has no component in $\mathrm{Span}(B)$ if $X^\dag y = 0$. Finding all the solutions to this linear equation thus gives us the subspace within $\mathrm{Span}(A)$ that is complementary to $\mathrm{Span}(B)$. By this procedure, we can obtain a reduced series of solutions $\{\tilde{\psi}_{R/Lj}^{(s;r)}\}$ where the redundancy due to the structure of Eq.~\eqref{eq:seqn_j} is removed. In doing so, we also confirm that no non-trivial solutions exist for $s\ge p$. Next, arranging the solutions into Jordan chains amounts to applying Eq.~\eqref{eq:jc} repeatedly along with finding the complement in the solution space to the constructed Jordan chains. For example, to start with, we may select one vector with the largest $s$ among all $j$, say $\tilde{\psi}_{Rj_0}^{(s_\mathrm{max};r=1)}$, then a Jordan chain of generalized right eigenvectors $\{\chi_{Rj_0}, \chi_{Rj_0+1}, \ldots, \chi_{Rj_0+s_\mathrm{max}-1}\}$ can be generated as
\begin{align}\label{eq:jceg}
    &\chi_{Rj_0} = \tilde{\psi}_{Rj_0}^{(s_\mathrm{max};r=1)},\;
    \chi_{Rj_0+\delta+1} = \mathbf{a}_{j_0+\delta+1,j_0+\delta} \chi_{Rj_0+\delta}, \nonumber\\
    &\qquad\qquad(\delta = 0,\ldots,s_\mathrm{max}-2).
\end{align}
After this Jordan chain is constructed, we can find its complement in the full solution space with all $j$ and $s$, and then repeat the above procedure in the complementary space. This process can be repeated until all solutions have been extracted, and, as a result, we obtain a complete set of Jordan chains.

In the example presented in Fig.~\ref{fig:generic_jcs}, we randomly generated the hopping amplitudes $a_{j+1,j;n}$s up to the third neighbors and the values are listed in the table below. The specific winding numbers chosen for this example are enforced by setting the amplitudes in each row to be maximum at a specific hopping range consistent with the winding number.
\begin{table}[h]
\begin{tabular}{@{}cccccccc@{}}
\toprule
\multirow{2}{*}{\begin{tabular}[c]{@{}c@{}}chiral flavors\\ ($j$+1, $j$)\end{tabular}} & \multicolumn{7}{c}{hopping range $n$} \\ \cmidrule(l){2-8} & $-3$   & $-2$   & $-1$   & $0$    & $1$    & $2$    & $3$    \\ \midrule
(2,1) & 0.41 & 0.54 & 0.70 & 1.04 & 0.78 & 1.53 & 0.39 \\
(3,2) & 0.33 & 0.55 & 0.76 & 1.05 & 0.71 & 1.49 & 0.38 \\
(4,3) & 0.30 & 0.53 & 1.69 & 1.00 & 0.65 & 0.59 & 0.35 \\
(1,4) & 1.41 & 0.50 & 0.76 & 1.02 & 0.77 & 0.47 & 0.31 \\ \bottomrule
\end{tabular}
\end{table}

\section{Boundary Charges when Reducing the Symmetry of Composite $p$}\label{sup:composite}

For a $p$CS system with composite $p = mn$, where $m,n\in \mathbb{N}\symbol{92} \{1\}$, we can break the $p$CS symmetry down into a reduced $m$CS. This means that the original $p$ chiral flavors are grouped such that we are left with $m$ flavors. In this appendix, we will see how to calculate the number of boundary charges for the new flavors in terms of the charges for the old flavors. We will only do the left boundary as the right boundary can be calculated in an analogous way.

We saw that breaking the symmetry into a smaller $m$CS leads to replacing the original vector invariant $\mathbf{W}$ of length $p$ to one of length $m$ with the components
\begin{equation}
    \left(\mathbf{W}^m\right)_r = \sum_{l=0}^{n-1}\left(\mathbf{W}\right)_{r+lm}.
\end{equation}
We consider only the left boundary, as the right boundary can be done in an analogous way. Following the recipe for calculating the boundary charges as outlined in the main text and derived in App.~\ref{sup:algo}, the new $\mathbf{Q}$ vector -- denoted here by $\mathbf{Q}^m$ -- is given by
\begin{equation}
    \begin{split}
        \left(\mathbf{Q}^m\right)_{s>1} &= \sum_{r=1}^{s-1}\left(\mathbf{W}^m\right)_r  = \sum_{r=1}^{s-1}\sum_{l=0}^{n-1}\left(\mathbf{W}\right)_{r+lm}\\
        &= \sum_{r=1}^{s-1}\sum_{l=0}^{n-1}\left[\left(\mathbf{Q}\right)_{r+1 + lm}-\left(\mathbf{Q}\right)_{r+ lm}\right]\\
        &= \sum_{r=1}^{s-1}\sum_{l=0}^{n-1}\left[\left(\mathbf{Q}_L\right)_{r+1 + lm}-\left(\mathbf{Q}_L\right)_{r+ lm}\right]\\
        &= \sum_{l=0}^{n-1}\left[\left(\mathbf{Q}_L\right)_{s + lm}-\left(\mathbf{Q}_L\right)_{1+ lm}\right],
    \end{split}
\end{equation}
where we have introduced the original $p$CS $\mathbf{Q}$ vector, and then rewritten the difference in terms of the corresponding left-boundary charge vector $\mathbf{Q}_L$. The new left-boundary charges are then $\mathbf{Q}^m_L = -\min(\mathbf{Q}^m) + \mathbf{Q}^m$, or more explicitly:
\begin{equation}
    \left(\mathbf{Q}^m_L\right)_s = -\min_s\left(\sum_{l=0}^{n-1}\left(\mathbf{Q}_L\right)_{s + lm}\right) + \sum_{l=0}^{n-1}\left(\mathbf{Q}_L\right)_{s + lm}.
\end{equation}
This implies that the left-boundary charges are simply given by first summing together the charges of all the old flavors that now belong the same new flavor, and then subtracting the smallest of these sums. This can be understood by remembering that we can move away charges from zero energy without breaking the symmetry only if we combine states corresponding to a complete set of flavors. This means that if we have all $m$ flavors at a boundary, we can remove as many charges from each flavor as there are complete $m$-tuplets. The number of complete $m$-tuplets equals the smallest charge number among the flavors.

\bibstyle{apsrev4-1}

%

\end{document}